\documentclass[10pt,twocolumn,letterpaper]{article}

\usepackage[pagenumbers]{cvpr}

\graphicspath{{figures/}}
\usepackage{amsmath}
\DeclareMathOperator*{\argmax}{\arg\!\max}

\usepackage{bbm}
\usepackage{bm}
\usepackage{caption}
\usepackage{subcaption}
\usepackage{algorithm2e}
\usepackage{algpseudocode}
\usepackage{cuted}
\usepackage{multirow}
\usepackage{float}
\usepackage[normalem]{ulem}

\definecolor{cvprblue}{rgb}{0.21,0.49,0.74}
\usepackage[pagebackref,breaklinks,colorlinks,allcolors=cvprblue]{hyperref}

\title{Kaleidoscopic Scintillation Event Imaging}

\author{
\vspace{1mm}
Alex Bocchieri$^1$ \quad John Mamish$^2$ \quad David Appleyard$^3$ \quad Andreas Velten$^{1}$ \\
\vspace{2mm}
{$^1$University of Wisconsin - Madison \quad $^2$Georgia Institute of Technology \quad $^3$Ubicept} \\
{\tt \small \url{https://bocchs.github.io/kaleidoscopic_scintillator}}
}

\begin{document}
\maketitle
\begin{abstract}
Scintillators are transparent materials that interact with high-energy particles and 
emit visible light as a result. 
They are used in state of the art methods of measuring high-energy particles and 
radiation sources.
Most existing methods use fast single-pixel detectors to detect and time 
scintillation events.
Cameras provide spatial resolution but can only capture an average over many events, 
making it difficult to image the events associated with an individual particle.
Emerging single-photon avalanche diode cameras combine speed and spatial resolution 
to enable capturing images of individual events.
This allows us to use machine vision techniques to analyze events, enabling new 
types of detectors.
The main challenge is the very low brightness of the events.
Techniques have to work with a very limited number of photons.

We propose a kaleidoscopic scintillator to increase light collection in a 
single-photon camera while preserving the event's spatial information.
The kaleidoscopic geometry creates mirror reflections of the event in known 
locations for a given event location that are captured by the camera.
We introduce theory for imaging an event in a kaleidoscopic scintillator
and an algorithm to estimate the event's 3D position.
We find that the kaleidoscopic scintillator design provides sufficient light 
collection to perform high-resolution event measurements for advanced radiation 
imaging techniques using a commercial CMOS single-photon camera.
\end{abstract}    
\section{Introduction}

\begin{figure}
\centering
\includegraphics[width=\linewidth]{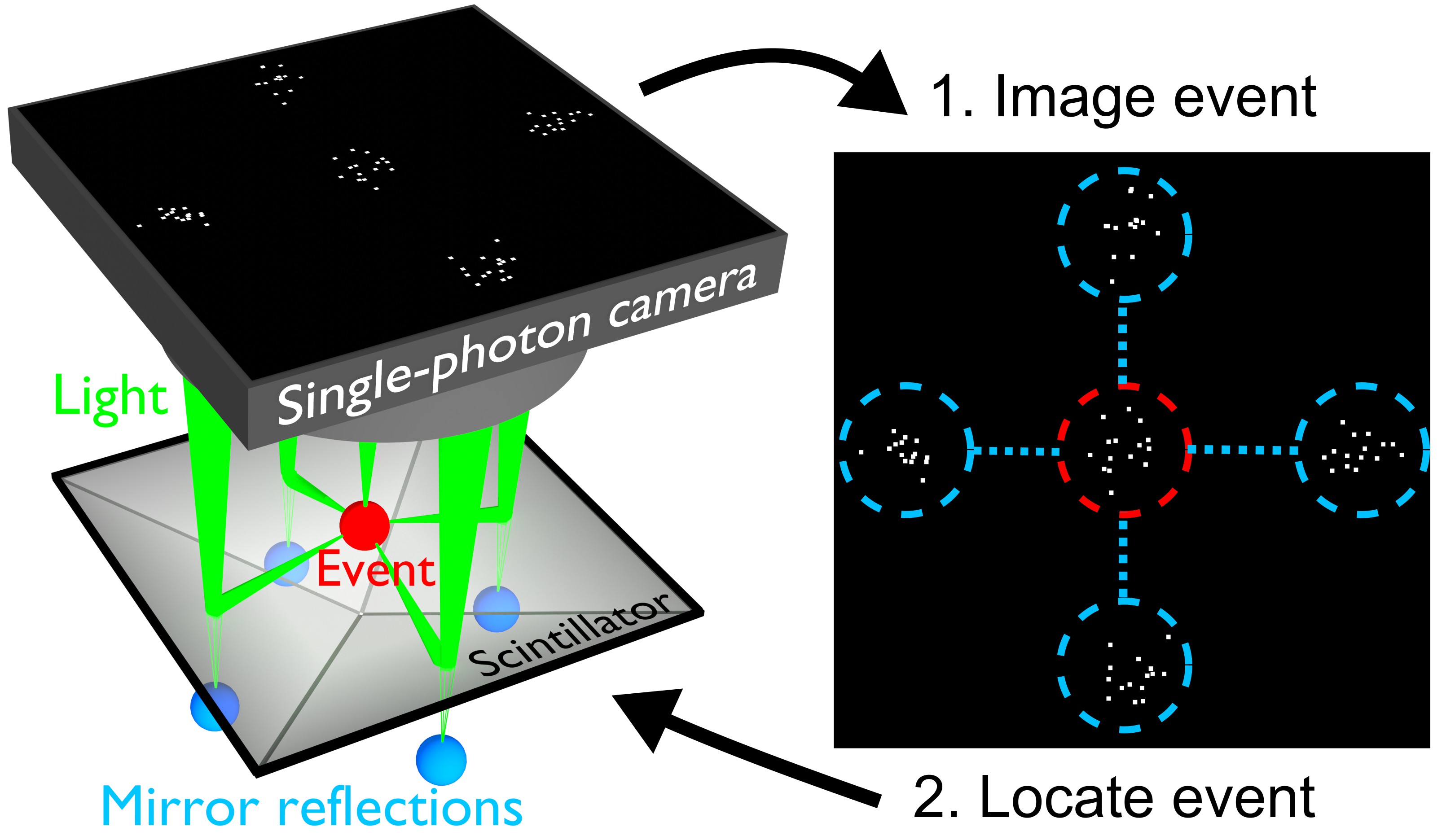}
\caption{\textbf{Method overview.}
An image is composed of light emitted from the scintillation event that 
reaches the camera either directly or after reflecting off mirrors of a 
kaleidoscopic scintillator.
The spatial relationship between the event and mirror reflections is embedded in a 
Gaussian mixture model whose likelihood is maximized to estimate the event's location using the EM algorithm.
} 
\label{fig:teaser}
\end{figure}
High energy particles are created in nuclear processes including from man-made 
devices and natural interactions. 
There is a wide variety of particles with varying properties. 
Detecting and characterizing them is crucial for a broad range of applications 
including nuclear security \cite{vetter2018gamma}, nuclear reactor and stockpile imaging \cite{beaumont2015high}, 
medical imaging \cite{gonzalez2021evolution}, 
archeology \cite{menichelli2007scintillating,ryzewski2013neutron}, 
and astronomy \cite{schonfelder1984imaging}. 
A particularly useful property of these particles is that many of them can 
penetrate through dense materials and therefore allow us to image through 
barriers and inside solid structures. 
The same penetrating properties, however, make it challenging to build a camera 
that can perform imaging or vision using high energy particles. 
A detector needs to have a large volume and density for a particle to interact. 
At the same time, we would like to know the location and shape of the interaction if it happens. 
One common approach to achieve this is by using a scintillator.

A scintillator is a transparent crystal that converts ionizing radiation into 
visible light.
It provides mass for an incident particle to be absorbed and detected with a photosensor.
While a particle is propagating through a scintillator, it may deposit energy and 
cause a ``scintillation event". 
In the case of gamma-ray radiation, a gamma-ray collides with an electron in the 
scintillator and causes the electron to recoil over a random walk.
A finite number of scintillation photons proportional to how much energy is 
deposited is emitted isotropically from the electron's path over a decay time.
The photons propagate out of the scintillator and are captured by a sensor.
Measurement of the event, such as its position, time, and energy deposition, is 
performed with the obtained signal.
These measurements are then used in various downstream tasks to characterize 
the radiation source.
Regardless of the task, optimizing light collection onto the sensor maximizes the 
signal to noise ratio (SNR) and event measurement performance 
(e.g. spatial, temporal, and energy resolutions).

This work focuses on optimizing imaging geometry in designs that estimate the 3D 
position of individual scintillation events with a single-photon camera.
Currently, such designs estimate an event's 3D position using depth from defocus 
and perspective projection \cite{bocchieri2024scintillation}.
These designs suffer from low light collection and are susceptible to noise from 
dark counts.
High-speed, single-photon sensors are required for tasks that measure individual 
events due to high particle incidence rates and the finite number of photons 
emitted by an event.
For these kinds of tasks, increasing exposure time does not increase light 
collection per event and decreases SNR, which makes maximizing light throughput crucial.
In general, large, thick scintillators are desirable because they increase the 
likelihood of events. 
However, estimating the 3D position of events in a larger volume is more challenging.
Applications where 3D event locations are measured include the Compton camera \cite{parajuli2022development}, 
neutron scatter camera \cite{weinfurther2018model}, 
positron emission tomography (PET), 
and single-photon emission computed tomography (SPECT) \cite{gonzalez2021evolution}.

We propose a kaleidoscopic scintillator geometry with specular surfaces to
increase light collection in a camera while preserving the event's spatial 
information.
A kaleidoscope consists of planar mirrors oriented like a pyramid frustum that 
contains the object being imaged.
Kaleidoscopes and light traps have been studied for imaging and reconstructing 
extended objects with abundant light \cite{reshetouski2011three,xu2018trapping}.
In this paper, we investigate the kaleidoscope as a technique for increasing light 
collection and reconstructing a point source in a photon-starved environment.
We study the kaleidoscope in the context of scintillation event imaging and pose a 
radiation detection problem as a computer vision problem.

The image of an event in a kaleidoscopic scintillator contains direct light from 
the event and indirect light from the event's mirror reflections.
Mirror reflections appear as events in locations determined by the event's 
location and the kaleidoscope's geometry, as illustrated in \cref{fig:teaser}.
The mirror reflections provide multiple views of the event and encode depth 
information with robustness to dark counts and low photon counts. 
Therefore, our proposed design improves event localization performance compared to 
previous non-kaleidoscopic methods.
We introduce a Gaussian mixture model (GMM) in which the spatial relationships 
between the event and mirror reflections are embedded and an algorithm that 
estimates the event's 3D location via maximum likelihood.
An overview of the method is shown in \cref{fig:teaser}.

Our contributions are summarized as follows: 
\begin{itemize}
\item A new scintillator design for increasing light collection in a camera while 
preserving the event's spatial information.
\item Theory for modeling light from a kaleidoscopic event that arrives at the 
camera sensor.
\item A probabilistic model of the image of a kaleidoscopic event with very few photons.
\item An algorithm to estimate an event's 3D location in a kaleidoscopic scintillator. 
The algorithm is validated on experimental data captured with a 
single-photon avalanche diode (SPAD) camera and a gamma-ray source. 
Improved 3D localization performance is demonstrated on experimentally-calibrated simulations.
\end{itemize}

\section{Related Work}

\noindent
\textbf{Kaleidoscopic vision.} Mirrors and kaleidoscopic designs have been used for stereo vision \cite{nene1998stereo, gluckman1999planar, gluckman2002rectified} 
and multi-view 3D reconstructions \cite{reshetouski2011three, ahn2021kaleidoscopic, ahn2023neural, kawahara2023teleidoscopic, fan2025kaleidoscopic, takahashi2021structure, bangay2004kaleidoscope, mitsumoto19923}
of an extended object using one picture from a single camera.
A ``light-trap" design consists of mirrors oriented such that light entering the 
trap reaches nearly every position inside the trap \cite{xu2018trapping}.
Time-of-flight is used to reconstruct a 3D object inside the trap.
In fact, the pyramid-shaped light-trap was found to provide the best object 
coverage, which is the same shape as the kaleidoscope we use in this work.

\noindent
\textbf{Scintillation event measurement designs.}
Numerous designs exist for measuring scintillation events.
Sensors such as silicon photomultiplier (SiPM) arrays or photomultiplier tubes 
(PMT's) can be coupled to the surface of a scintillator that is monolithic or 
pixelated with internal waveguides \cite{gonzalez2021evolution,kawula2021sub,hosokoshi2019development,lee2020development,kataoka2013handy}.
Camera designs measure events in a monolithic scintillator by imaging the 
scintillator with a lens \cite{bocchieri2024scintillation,losko2021new,gustschin2024event,yamamoto2023development,d2021novel,gao2023novel,pleinert1997design,baker2014scintillator,adams2017gamma,balasubramanian2022x}.
Using a camera to estimate the 2D position of an event in a thin scintillator, 
not including depth, is performed using center of mass algorithms \cite{hussey2017neutron,losko2021new,gustschin2024event}.

There is limited work that estimates the 3D position of individual events with a 
camera. 
One work attempts to remove dark counts and estimate depth from defocus \cite{bocchieri2024scintillation}.
Another work uses stereo vision with two objectives and mirrors inside the imaging 
system to project two views of a cubic scintillator onto a hybrid photon detector \cite{filipenko2014three}.
Our proposed system provides multiple views, is more efficient, and uses a 
commercially available CMOS single-photon camera.
A plenoptic SPAD camera was proposed to reconstruct events \cite{dieminger2025ultrafast}.
The plenoptic camera design's experimental demonstration was 
limited by low light collection, obtaining up to 4 photons in an image,
while its simulated reconstruction of neutrino tracks achieved high resolution.

\section{Kaleidoscopic Event Imaging Theory} \label{sec:theory}

We consider a square pyramid scintillator throughout this paper. 
Each face of the scintillator, except for the base, is a specular surface.
The scintillator has an index of refraction $n>1$ and height $h$.

\subsection{Imaging Configuration and Model}

\begin{figure}
\centering
\includegraphics[width=\linewidth]{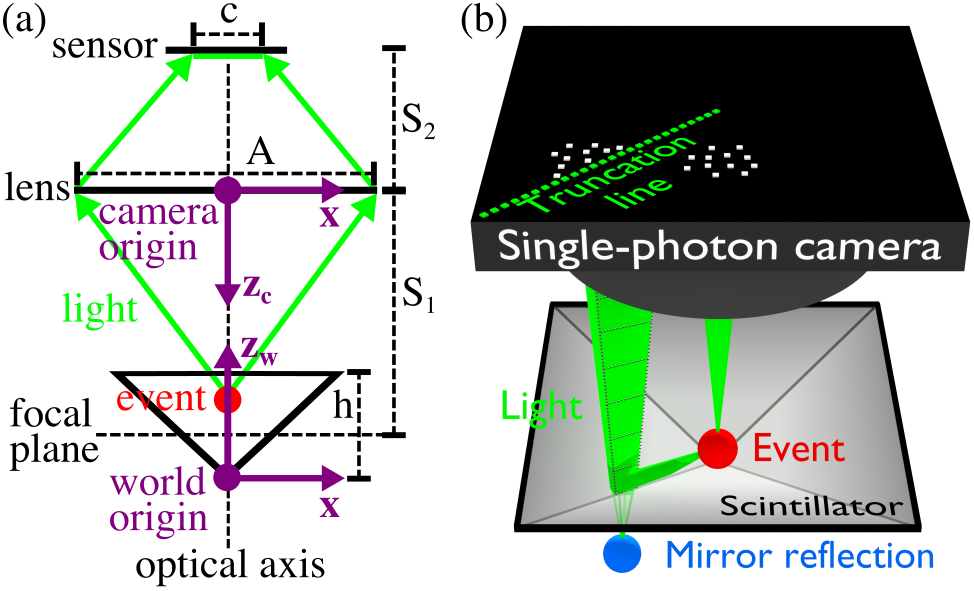}
\caption{\textbf{(a) Imaging parameters and coordinate systems.}
Only light emitted directly from the event to the camera is shown.
\textbf{(b) Image truncation example in 3D.}
The image of an event and one mirror reflection is shown. 
Light corresponding to the mirror reflection is truncated at the mirror's edge, 
resulting in a truncation line in the image.
The truncation line only applies to that mirror reflection.}
\label{fig:trunc_plus_config}
\end{figure}

The world coordinate system's origin is set at the pyramid's apex with the 
$z$-axis directed perpendicular toward the pyramid's base surface.
The camera coordinate system's origin is at the center of the lens with its 
$z$-axis directed toward the scintillator in the opposite direction of the world's $z$-axis.
The $x$ and $y$ axes among the two coordinate systems are in the same direction.
We denote a $z$-coordinate in the world coordinate system as $z_w$ and camera 
coordinate system as $z_c$.
Transforming between the two coordinate systems is done by subtracting the 
$z$-coordinate from the lens' $z$-coordinate in the world coordinate system.
We use a thin lens to model the camera.
Three planes to note are the focal plane, thin lens plane, and sensor plane.
The focal plane is set at the apparent depth of the pyramid's apex at $z_w=h-h/n$.
A thin lens with diameter $A$ is placed at a distance $S_1$ from the focal plane. 
The sensor is placed at a distance $S_2$ from the lens.
The lens' focal length is set to $f=(S_1^{-1}+S_2^{-1})^{-1}$.
Parameters are illustrated in \cref{fig:trunc_plus_config}a.

Throughout this paper, the scintillator is oriented so that its base surface's 
edges are parallel with the sensor's respective edges. 
In this manner, the $x$-$y$ dimensions of the normal vector for each specular 
surface of the scintillator are aligned along either the $x$ or $y$ axes.
We denote the scintillator's four specular surfaces as +$x$, +$y$, -$x$, or -$y$ mirrors.

A scintillation event is approximated as a point source of light emitting in all directions.
Since the optical setup is constrained to short imaging distances, 
the images of an event and its mirror reflections exhibit defocus blur and 
have nonzero diameters.
An image's diameter on the sensor varies according to the event's distance from 
the focal plane, following the circle of confusion model.
Note that the camera sees an event at its apparent depth rather than its real 
depth due to the scintillator's index of refraction.
An event's real and apparent locations are separated by $d-d/n$ along the 
$z$-dimension, where $d$ is the event's real depth in the scintillator from the
base surface.
We denote an apparent location using the superscript ``$(a)$" and a real location 
by the lack of a superscript.
For an event at an apparent location $(x_0,y_0,z_{c0}^{(a)})$, 
the circle of confusion model yields
\begin{equation} \label{eqn:circ_of_conf}
c=A\frac{S_2}{S_1}\frac{|S_1-z_{c0}^{(a)}|}{z_{c0}^{(a)}}
\end{equation}
where $c$ is the image diameter at the sensor.

\begin{figure*}
\centering
\includegraphics[width=\linewidth]{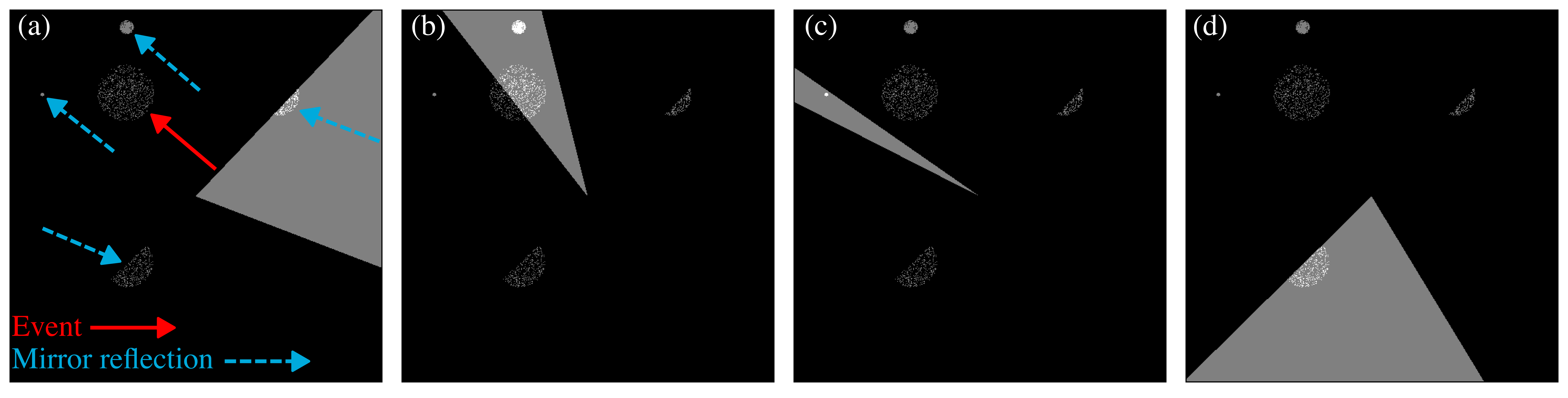}
\caption{\textbf{Simulated kaleidoscopic image with theoretical acceptance zones.} 
Acceptance zones are derived for each mirror reflection and overlaid in gray on 
the image for the (a) +$x$, (b) +$y$, (c) -$x$, and (d) -$y$ mirror reflections.} 
\label{fig:trunc_examples}
\end{figure*}

\subsection{Mirror Reflections and Apertures}
Consider the real location of an event at $\bm{p_0}$.
Mirror $k$ produces a mirror reflection with a real location at 
$\bm{p_k}=\bm{T_k}\bm{p_0}$
where
\begin{equation} \label{eqn:ref_trans}
\bm{T_k}=\bm{I}_{3\times3} - 2\bm{n_k}\bm{n_k}^T
\end{equation}
is the mirror's transformation, $\bm{I}$ is the identity matrix, and $\bm{n_k}$ is 
the mirror's normal vector.
The mirror reflection of an event is also a point source of light.
The captured image of the mirror reflection is obtained from the photons that 
reflect off the mirror and into the camera, exhibiting the same defocus blur as if 
an event were located at $\bm{p_k}$.
However, due to the finite mirror size, the image on the sensor may be truncated 
along lines corresponding to the mirror's edges.
This occurs when $\bm{p_k}$ is behind another mirror from the camera's perspective.
The photons that are truncated from the image are those that reflect off the 
mirror adjacent to mirror $k$ near the shared edge, as shown in \cref{fig:trunc_plus_config}b.
Essentially, mirror $k$ behaves like an aperture to a light source at $\bm{p_k}$.
We denote the area on the sensor where photons from a mirror reflection cannot 
arrive as the ``truncation zone" and its complement as the ``acceptance zone".
A line that separates the truncation zone and acceptance zone is a ``truncation line".
A 3D visualization of light truncation is shown in \cref{fig:trunc_plus_config}b.
Supplementary \cref*{fig:trunc_theory} shows a 2D view of the propagation of 
light and how truncations form in various scenarios.
Truncation lines are derived in \cref*{sec:image_truncations_sup}.

Light that reflects over multiple mirrors may be stopped by any of those mirrors' 
edges and also cause image truncations.
We term a mirror reflection's ``order" as the number of reflections its light 
underwent before forming. 
Light reflecting off mirror $k$ can reflect off another mirror $l$ and generate a 
mirror reflection at $\bm{p_l}=\bm{T_l} \bm{p_k}$.
The light incident on mirror $l$ from $\bm{p_k}$ passes through the aperture 
of mirror $k$.
If this light spans a partial area $A_l$ of mirror $l$, then mirror $l$'s aperture 
is $A_l$.
Otherwise, mirror $l$'s aperture is simply the mirror.
Mirror $l$'s aperture affects the light emitted from $\bm{p_l}$ toward another 
mirror for a higher-order reflection and toward the camera for imaging.
Higher-order reflections and imaging continue in the same manner.

\subsection{Imaging Theory Validation}
We validate the theory on mirror reflections and image truncations by observing 
that theoretically derived truncation lines align with photon arrivals in a 
simulated image.
We simulate the image of an event in a kaleidoscopic scintillator in the shape of 
a square pyramid using a thin lens and ray tracing. 
We simulate an unrealistically high number of photons so that image truncations 
are clearly observable.
Simulation details are in \cref*{sec:sim_details}.
Acceptance zones for each mirror reflection are derived as in 
\cref*{sec:image_truncations_sup} and overlaid on the image.
The resulting image is shown in \cref{fig:trunc_examples}.
\section{Event Localization Algorithm}

\subsection{Image Model}

We adopt a 2D Gaussian 
\begin{equation}
\mathcal{N}(\bm{t};\bm{\mu},\sigma^2)=\frac{1}{2\pi\sigma^2}\exp\left({-\frac{||\bm{t}-\bm{\mu}||_2^2}{2\sigma^2}}\right) 
\end{equation}
as the camera's point spread function and assume a circular Gaussian with 
covariance matrix $\bm{\Sigma}=\sigma^2 \bm{I}_{2\times2}$. 
$\sigma=ac$, where $a$ is a proportionality constant determined by the 
optical configuration, and $c$ is the circle of confusion diameter in \cref{eqn:circ_of_conf}.
$\bm{t}$ is a 2D coordinate on the sensor plane.

An event or mirror reflection at apparent location 
$(x_k,y_k,z_{ck}^{(a)})$ is modeled as 
a point source of light, so its image on the sensor consists of photon arrivals 
spatially distributed over a 2D Gaussian with mean 
\begin{equation} \label{eqn:mu}
\bm{\mu_k}=\left[ \frac{S_2}{z_{ck}^{(a)}}x_k, \frac{S_2}{z_{ck}^{(a)}}y_k \right]
\end{equation}
and standard deviation 
\begin{equation} \label{eqn:stdev}
\sigma_k=aA\frac{S_2}{S_1}\frac{|S_1-z_{ck}^{(a)}|}{z_{ck}^{(a)}}
\end{equation}
where \cref{eqn:mu} is derived from perspective projection.

We model the image of an event in a kaleidoscopic scintillator with $K$ mirror 
reflections and $N$ photons as a GMM.
The GMM has $K+1$ components that correspond to the event or a mirror reflection, 
and each photon in the image is a sample of the GMM.
The location of each mirror reflection at $\bm{p_k}$ for $k=1...K$ generated from 
an event at $\bm{p_0}=(x_0,y_0,z_{w0})$ is known based on the kaleidoscope's geometry.
For any mirror reflection $k$, each coordinate in $(x_k,y_k,z_{wk})$ is a linear 
combination of the event's $(x_0,y_0,z_{w0})$ coordinates based on the mirror's 
reflection transformation.
All $\bm{p_k}$'s can be written in terms of $\bm{p_0}$ using 
\cref{eqn:ref_trans}, and each Gaussian component's $\bm{\mu_k}$ and $\sigma_k$
can be written in terms of $\bm{p_0}$ by transforming to camera coordinates and using \cref{eqn:mu,eqn:stdev}.
Thus, every Gaussian component is constrained to $\bm{p_0}$.
This results in an optimization problem for estimating $\bm{p_0}$ that captures 
the global information of the event and all mirror reflections:
\begin{equation} \label{eqn:opti_prob}
\argmax_{x_0,y_0,z_{w0},\bm{\pi}} Q - \lambda \sum_{k=0}^K \left \lVert \bm{\mu_k} - \bm{\mu_k}^0 \right \rVert_2^2 
\end{equation}
where 

\begin{equation}
\begin{split}
Q = \sum_i \sum_k r_{ik} \bigg[ \text{log}(\pi_k) + \text{log}(w_i) - \text{log}(2\pi{\sigma_k}^2) \\ - \frac{w_i}{2{\sigma_k}^2}||\bm{t_i}-\bm{\mu_k}||_2^2 \bigg]
\end{split}
\end{equation}

\begin{align} \label{eqn:r_ik}
r_{ik} = \frac{\pi_k \mathcal{N}(\bm{t_i};\bm{\mu_k},{\sigma_k^2})}{\sum_{k'=0}^K \pi_{k'} \mathcal{N}(\bm{t_i};\bm{\mu_{k'}},{\sigma_{k'}^2})}
\end{align}
$Q$ is the expected value of the weighted complete-data log-likelihood,
$r_{ik}$ is the probability of photon $i$ belonging to component $k$ given current 
parameter values, 
$w_i$ is the weight assigned to photon $i$, 
$\pi_k$ is the mixing weight for component $k$, 
$\bm{\mu_k}^0$ is the initialization point for $\bm{\mu_k}$ as described in the 
initialization procedure below, 
and $\lambda$ is a regularization coefficient.
We adopt a density-based weighting scheme for photon samples to minimize the 
influence of sparsely distributed dark counts and define
$w_i = \sum_{j \in S_i^q} \text{exp} \left( -\frac{||\bm{t_i}-\bm{t_j}||_2^2}{\nu} \right)$,
where $S_i^q$ is the set of $q$ nearest neighbors of photon $i$, and $\nu$ is a 
positive scalar.
We use the EM algorithm to optimize \cref{eqn:opti_prob}.

The above GMM formulation and definitions are derived in \cref*{sec:image_model_sup}.
$\bm{p_k}$ is written in terms of $\bm{p_0}$ in \cref*{sec:like}.

\subsection{Optimization Algorithm}
We assume a square pyramid kaleidoscope geometry, up to first-order reflections, 
and the presence of at least two mirror reflections in an image.
Each edge of the scintillator's square surface is parallel to each respective edge 
of the sensor.
In this configuration, mirror reflections' $x$-$y$ coordinates will be located 
along either the $\pm x$ or $\pm y$ directions from the event's location.

One or more mirror reflections may be missing from the image due to truncations 
depending on the event's location.
Therefore, determining which mirror reflections are present is required to compute \cref{eqn:opti_prob}.
We run the following initialization procedure to determine the presence of mirror 
reflections and to initialize the event's estimated location for the EM algorithm.

\noindent
\textbf{Initialization procedure.}
The initialization procedure can be summarized as maximizing \cref{eqn:opti_prob}
over a set of possible event locations and $C \in \{3,4,5\}$ clusters 
(number of mirror reflections plus the event).
First, centroids are computed using weighted KMeans with $C$ clusters and the 
photons' assigned weights.
Centroids are then classified as either the event, or $+x$, $-x$, $+y$, or $-y$ 
mirror reflections based on their relative positioning.
This is done by taking combinations of subsets of centroids and computing the 
standard deviation of a subset's coordinates along the $x$ and $y$ dimensions to 
determine which centroids are horizontally or vertically aligned.
We denote these centroids as $\bm{\mu_k}^0$.
A set of possible event locations that is uniformly spaced over the depth 
($z$ dimension) of the scintillator is computed using the event's centroid and \cref{eqn:mu}.
The number of mirror reflections, which reflections are present, and the 
initialization point for $\bm{p_0}$ are those that correspond to the 
highest value of \cref{eqn:opti_prob} out of this set and over $C \in \{3, 4, 5\}$.
When computing \cref{eqn:opti_prob}, terms that correspond to a mirror reflection 
$k$ are included only if that mirror reflection is determined to be present.
$\bm{\pi}$ is initialized to the uniform distribution.

\noindent
\textbf{Optimization procedure.}
During the E-step, $r_{ik}$ is updated using \cref{eqn:r_ik} and 
the current values of $\bm{p_0}$ and $\bm{\pi}$.
If photon $i$ is distant from any $\bm{\mu_k}$ and computing the denominator in 
\cref{eqn:r_ik} results in underflow, then we set $r_{i:}=0$ and do 
not use photon $i$ in computations.
During the M-step, $\bm{p_0}$ is updated by fixing $\bm{\pi}$ and optimizing 
\cref{eqn:opti_prob} with gradient ascent. 
The gradient with respect to $\bm{p_0}$ is derived in \cref*{sec:grad}.
$\bm{\pi}$ is then updated using $\pi_k=\frac{1}{N}\sum_{i=1}^N r_{ik}$.
In both the E and M steps, terms that correspond to a mirror reflection $k$ are 
included in the computation only if that mirror reflection is present in the image.

\noindent
\textbf{Regularization.}
We introduce the regularization term $\lambda \sum_{k=0}^K \left \lVert \bm{\mu_k} - \bm{\mu_k}^0 \right \rVert_2^2$ 
to favor solutions where all components' $\bm{\mu}$'s are close to their 
corresponding $\bm{\mu}^0$.
This regularization is most helpful in cases where the mirror reflection photon 
clusters are close to each other.
Without regularization, the algorithm may find an optimum that groups the event 
and one or more mirror reflections together into one component with a large 
$\sigma$ while having other components' $\bm{\mu}$'s that do not coincide with a 
photon cluster.
An example of this is shown in Supplementary \cref*{fig:regularization}.

\section{Hardware Experiments} \label{sec:exp}

\begin{figure}
\centering
\includegraphics[width=\linewidth]{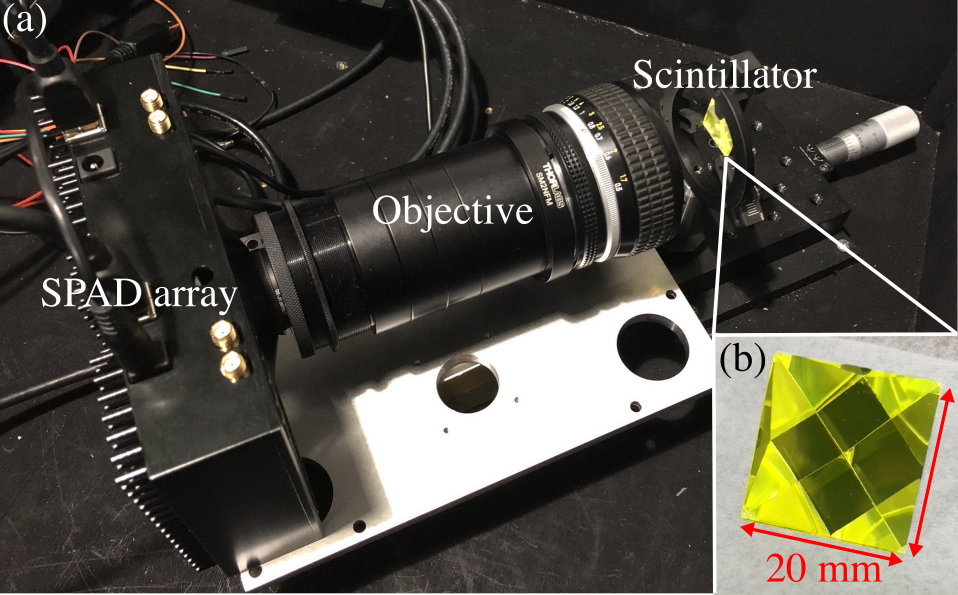}
\caption{\textbf{Experimental setup.} (a) The setup without the gamma-ray source. (b) The kaleidoscopic scintillator.}
\label{fig:setup}
\end{figure}

\begin{figure*}
\centering
\includegraphics[width=\linewidth]{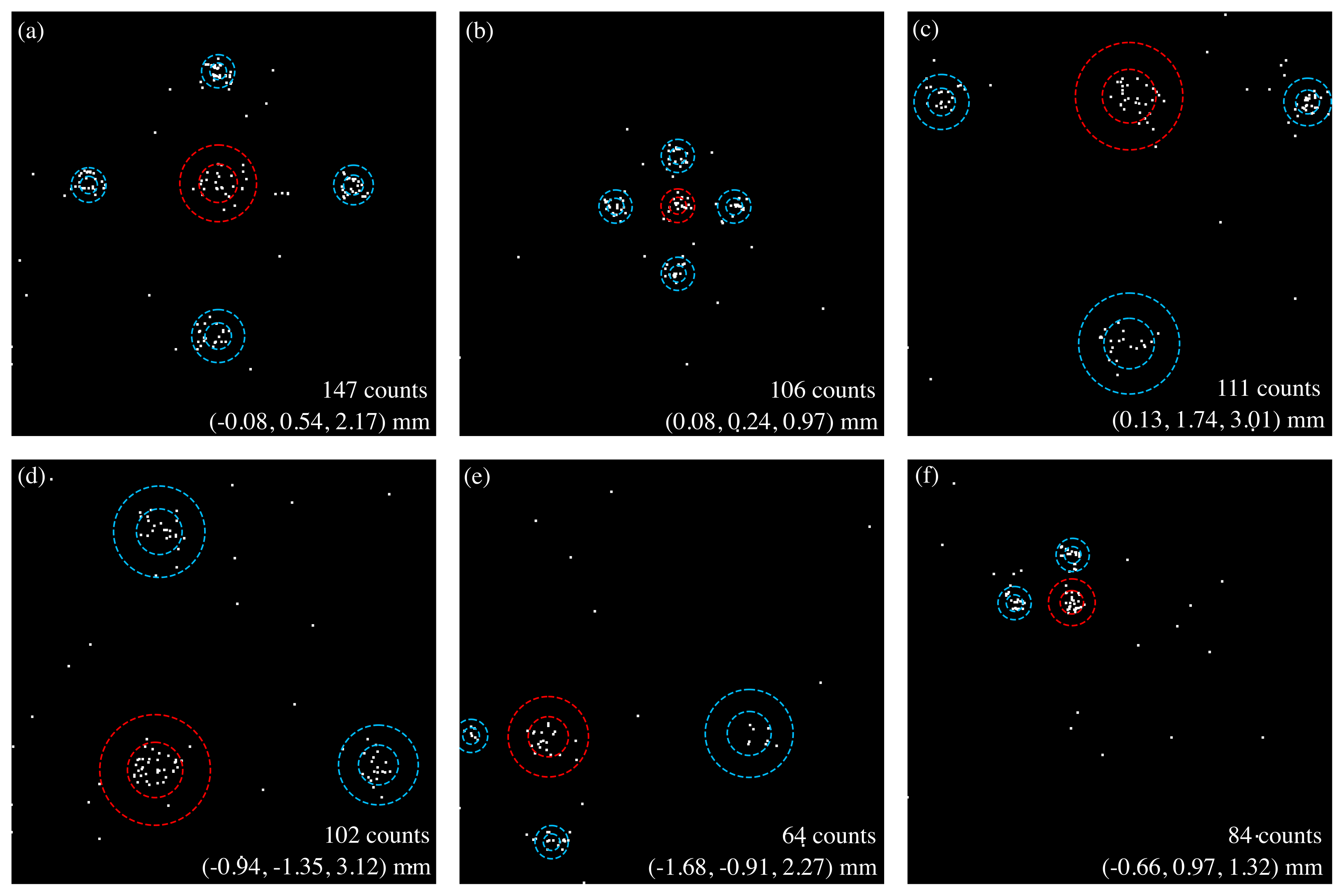}
\caption{\textbf{Selected experimental images.} Experimental images overlaid with the algorithm's estimated Gaussian components.
Each dashed circle is centered on the Gaussian component's mean. 
The inner and outer circles are one and two standard deviations in radius, respectively.
Red and blue circles represent the event and mirror reflections, respectively.
Pixels with a photon are enlarged with a $3 \times 3$ filter for visualization purposes.
The number of counts in the image and the algorithm's estimated event location are
shown in each image.
} 
\label{fig:example_figures}
\end{figure*}

\noindent
\textbf{Hardware and data collection.}
The experimental hardware consists of a SPAD array, lens, scintillator, and 1 $\mu$Ci Co-60 gamma-ray source (1.17, 1.33 MeV).
We use the SPAD512 (Pi Imaging) array with microlenses for increased fill factor, 
which has $512 \times 512$ pixels and 16 $\mu$m pixel pitch.
The lens is a 50 mm focal length Nikkor lens set to a f/1.2 aperture.
We use a GAGG(Ce)-HL scintillator (Epic Crystal), which has a 150 ns decay 
constant, a 530 nm emission peak, and an index of refraction of 1.91.
Its geometry is a square pyramid with a 20 mm wide base, 5.77 mm 
height, and a 120 degree opening angle at the apex.
This wide angle is chosen so that second-order reflections are unlikely to occur 
or be imaged.
Four surfaces of the scintillator are coated with enhanced specular reflector.
The SPAD array is configured to capture 1-bit images with 1.5 $\mu$s 
integration time to minimize the accumulation of dark counts while 
allowing time for scintillation light to be emitted.
The camera's lateral field of view (FOV) covers approximately $5.5 \times 5.5$ mm 
at the focal plane.
The scintillator is positioned such that its apex is in-focus and centered in the 
camera's FOV, and that the scintillator's base edges are parallel with the 
sensor's respective edges.
An air gap of approximately 30 mm exists between the scintillator and the lens.
The entire setup is placed inside a light-tight enclosure to keep ambient light out.
The experimental setup and scintillator are shown in \cref{fig:setup}.
The experimental camera focus is shown in Supplementary \cref*{fig:experiment_focus}. 

Data collection took place at about 21 degrees Celsius ambient temperature.
We first captured 130,000 images without the gamma-ray source to compute a mask 
of 5\% of pixels with the highest dark count rates. 
We zero the pixels in this mask in all experimental images.
After zeroing 5\% of pixels, we observe a median of 4 dark counts per image.
A histogram of dark counts per image is shown in Supplementary \cref*{fig:dark_counts_hist}.

We collect data with the the gamma-ray source placed adjacent to the 
scintillator's apex and passively capture images since the timing of gamma-ray 
emissions cannot be controlled.
All computations are performed in post-processing.
We capture 13,000,000 images with the gamma-ray source present and 
discard images with less than 60 counts, resulting in 4,379 images.
Histograms of counts in an image are shown in Supplementary \cref*{fig:cap_counts_hist}.

\noindent
\textbf{Parameter values.}
In all experiments, we set $\lambda=10$.
We clip all $\sigma_k$'s to a a minimum value of 10 pixels to account for 
imperfect focusing and a non-ideal point source, and to prevent singularities.
The gradient of $\sigma_k$ is set to 0 if $\sigma_k$ gets clipped to the minimum value.
See \cref*{sec:param_values} for the values of all the 
algorithm's parameters and \cref*{sec:param_cal} for how camera 
parameters are calibrated.

\noindent
\textbf{Kaleidoscopic image model validation.}
We select six experimental images and overlay the algorithm's estimated Gaussian 
components, shown in \cref{fig:example_figures}, to validate the presence of 
mirror reflections in accordance with the kaleidoscopic model.
Additional experimental images are shown in Supplementary \cref*{fig:example_figures_sup}.

\noindent
\textbf{Localization algorithm cross-validation.}
Experimental events cannot be controlled, so their ground truth locations are unknown.
Therefore, to validate that the algorithm is measuring the event's location, we 
report agreement of multiple measurements of the event's location as follows.
We use experimental images that contain the event and four mirror reflections as 
test images.
The number of mirror reflections in an image is determined using the algorithm's 
initialization procedure.
Then, we create new images of the event by removing combinations of one or two 
mirror reflections from the image.
Photon $i$ is classified as belonging to mirror reflection $k$ according to 
$\max_k r_{ik}$ using $r_{ik}$ values obtained after running the algorithm's 
optimization procedure on the original test image with four mirror reflections.
Thus, for one test image, we generate 4 images with one mirror reflection removed 
and 6 images with two mirror reflections removed for a total of 11 images 
corresponding to one event.
We run the algorithm's initialization and optimization procedures on each image to 
obtain 11 measurements of the event's location.
We compute the mean estimated event location over the 11 images corresponding to 
one event.
We record the distance between the mean location and the estimated location for 
each individual image.
The distribution of this distance over all images is used to report the 
agreement in event location measurements, where short distances indicate good 
agreement.
This metric also provides a measure of experimental precision.
Selecting images by using the algorithm that contain four mirror reflections and 
at least 60 counts resulted in 1,606 test images.
The median, mean, and standard deviation, respectively, of 
distances are 0.10 mm, 0.17 mm, and 0.22 mm over 17,666 distances.
Distances are reported in a histogram in Supplementary \cref*{fig:crossval_error}.

To confirm that mirror reflections are being correctly identified and removed, 
we record the fraction of photon counts in an image that are removed.
We assume that an event and each mirror reflection have the same average brightness.
Therefore, we expect about 1/5 and 2/5 of counts in an image to be removed when 
removing one and two mirror reflections, respectively.
The median, mean, and standard deviation, respectively, of the fraction of counts 
removed in an image are 0.19, 0.18, and 0.07 over 6,424 images where one mirror 
reflection has been removed, and 0.38, 0.37, and 0.08 over 9,636 images where two 
mirror reflections have been removed.
The distribution of fractions of counts removed are reported in histograms in 
Supplementary \cref*{fig:removal_frac}.

\noindent
\textbf{Regularization ablation study.}
We perform an ablation study to demonstrate the effectiveness of regularization 
in \cref{eqn:opti_prob}.
Details are in \cref*{sec:ablation}.

\section{Simulations} \label{sec:simulations}

\noindent
\textbf{Comparison to prior art}.
Since scintillation events are difficult to produce at controlled locations, 
direct evaluation of the method requires simulated data.
Here, we simulate our system and compare its localization performance to prior 
3D localization methods, which estimate depth from defocus in the absence of 
mirror reflections \cite{bocchieri2024scintillation}.
We evaluate position accuracy and resolution over a grid of event locations and 
over different levels of event brightness.
We use the same scintillator geometry, index of refraction, calibrated camera 
parameter values, and sensor array specifications as in the experiments.

\begin{table}
\footnotesize
\centering
\begin{tabular}{cccccc} 
 \toprule
 & & \multicolumn{4}{c}{Average error and resolution (mm) $\downarrow$} \\
 \cmidrule(lr){3-6}
 $N_0$ & Algorithm & 3D Error & x Res. & y Res. & z Res. \\
 \midrule
  \multirow{3}{*}{30} & Kaleid. (ours) & \textbf{0.14} & \textbf{0.16} & \textbf{0.14} & \textbf{0.14}  \\
  & Non-kaleid. (ours) & 0.83 & 1.36 & 1.41 & 1.32  \\
  & Noise-removing \cite{bocchieri2024scintillation} & 0.64 & 1.04 & 1.02 & 1.04 \\
 
 \midrule
  \multirow{3}{*}{20} & Kaleid. (ours) & \textbf{0.16} & \textbf{0.15} & \textbf{0.17} & \textbf{0.16} \\
  & Non-kaleid. (ours) & 0.79 & 1.17 & 1.24 & 1.21 \\
  & Noise-removing \cite{bocchieri2024scintillation}  & 0.68 & 1.05 & 1.03 & 1.06   \\
 
 \midrule
 \multirow{3}{*}{10} & Kaleid. (ours) & \textbf{0.29} & \textbf{0.27} & \textbf{0.31} & \textbf{0.27}  \\
  & Non-kaleid. (ours) & 0.95 & 1.36 & 1.33  & 1.28 \\
  &  Noise-removing \cite{bocchieri2024scintillation} & 0.80 & 1.20 & 1.13 & 1.14  \\

 \bottomrule
\end{tabular}
\caption{\textbf{Simulation results.}}
\label{tab:sim_results}
\end{table}

\noindent
\textbf{Image generation.}  
For a given event location, we generate 100 kaleidoscopic images for each of three 
brightness levels as follows.
We compute the locations of the mirror reflection and draw $Poisson(N_0)$ photons 
from each Gaussian component parameterized according to \cref{eqn:mu,eqn:stdev} 
and accounting for the index of refraction.
As in the experiments, all $\sigma_k$'s are clipped to a minimum value of 10 
pixels, which corresponds to $z_w=0.82$ mm with the given experimental parameter values.
Acceptance zones are computed, as described in \cref*{sec:image_truncations_sup},
and photons that lie outside their mirror reflection's acceptance zone are removed.
We test brightness levels over $N_0 \in \{10, 20, 30\}$.
$Poisson(10)$ dark counts are uniformly randomly added over each image.
At each location, we also generate 100 non-kaleidoscopic images in the same 
manner, except that no mirror reflections are generated.

\noindent
\textbf{Event locations.}
The grid of event locations spans $x \in (0,2.5)$ mm, $y \in (0,2.5)$ mm, and
$z_w \in (0.82, 3.5)$ mm.
This grid is selected so that event locations span one $x$-$y$ quadrant due to 
the system's symmetry, and $\sigma_0$ corresponding to the event's photon 
cluster is above the minimum clipping threshold.
Grid points are equispaced over 10 points in each dimension. 
If less than two mirror reflections' centroids ($\bm{\mu_k}$'s) are in their 
acceptance zones and within the sensor, then the event location is discarded.
These discarded locations correspond to points near the edge of the camera's FOV, 
and we consider their resulting image to be inadequate for applying our proposed algorithm.
Event locations that lie outside the scintillator are also discarded.
This results in a total of 463 valid event locations.

\noindent
\textbf{Algorithms.}
The proposed algorithm is tested on the kaleidoscopic images with the same 
parameter values and initialization and optimization procedures as on the 
experimental data.
A non-kaleidoscopic version of the proposed algorithm is tested on 
non-kaleidoscopic images.
The non-kaleidoscopic version of the algorithm is the same as the kaleidoscopic 
version, except that initialization points are lower-bounded at $z_w=0.82$ mm, 
only one cluster or Gaussian component is computed, 
and $r_{ik}=\mathbbm{1}\{k=0\}$ for all photons.

We also test a prior method on non-kaleidoscopic images that attempts to classify 
and remove dark counts and then directly solve for the event location's maximum 
likelihood estimate without weighting photons \cite{bocchieri2024scintillation}.
More details of this method are in \cref*{sec:prior_method}.
The algorithms that are applied to non-kaleidoscopic images estimate depth from defocus.

\noindent
\textbf{Performance metrics.}
At each location, we compute the mean 3D error (Euclidean distance) over all 
images and the spatial resolution.
Spatial resolution is defined as 2.355$\sigma_e$ (full width half maximum), 
where $\sigma_e$ is the standard deviation of one dimension of the estimates of an 
event location.
We report the averages of the mean 3D error and the spatial resolution in each 
dimension taken over all event locations for each brightness level 
$N_0 \in \{10, 20, 30\}$ for the three different algorithms. 
\cref{tab:sim_results} contains results for the kaleidoscopic and 
non-kaleidoscopic versions of the proposed algorithm and for the prior method with 
$T_{edge}=80$ pixels (see \cref*{sec:prior_method} for details).
The prior method's results using different $T_{edge}$ values are reported in
Supplementary \cref*{tab:sim_results_supp}.

\section{Discussion}

\noindent
\textbf{Simulated validation of theoretical image truncations.}
The mirror reflection images generated with ray tracing and a thin lens are 
truncated along the theoretically derived acceptance zones, shown in \cref{fig:trunc_examples}. 
This suggests the theoretical models for mirror reflection locations and 
image truncations are correct.
In \cref{fig:trunc_examples}, the +$x$ and -$y$ mirror reflections are 
partially truncated, while the +$y$ and -$x$ mirror reflections are not truncated.

\noindent
\textbf{Experimental validation of kaleidoscopic image model.}
Various examples where the Gaussian components predicted by the 
algorithm coincide with the photon clusters of the event and mirror 
reflections are shown in \cref{fig:example_figures}.
This demonstrates that mirror reflections are indeed being captured in accordance 
with the proposed model.
Examples include events occurring at high (\cref{fig:example_figures}a) and low 
(\cref{fig:example_figures}b) $z_w$-coordinates, mirror reflections lying outside 
the FOV (\cref{fig:example_figures}c,d), and mirror reflections that are 
completely truncated (\cref{fig:example_figures}e,f).

Dark counts in \cref{fig:example_figures} are higher than the median dark count 
rate of 4 pixels per image without the gamma-ray source present.
This could be due to several factors, including increased cross talk from higher 
photon collection, fluorescence from events from other particles or gamma-rays 
occurring before the beginning or at the end of an image's exposure time, 
imperfect mirror reflections, or internal reflections.

\noindent
\textbf{Experimental cross-validation of localization algorithm.}
The distances between the mean and individual estimated event locations over 
images with removed mirror reflections are small, demonstrating an agreement in 
event location measurements among different images of the same event with removed 
mirror reflections.
This agreement indicates the algorithm is in fact measuring the event's 
location and not spurious light sources.
The distributions of fraction of photon counts removed from an image when removing 
one and two mirror reflections are centered close to 1/5 and 2/5, indicating that
photons from a mirror reflection are being classified correctly by the algorithm.
These results would be unlikely if the algorithm were not measuring the event, or 
if images were of spurious light that does not originate from an event or follow 
the kaleidoscopic model.
Further proof that the algorithm is measuring events is seen in simulations where 
the ground truth is known.

\noindent
\textbf{Simulations.}
The results in \cref{tab:sim_results}
show that the kaleidoscopic design improves 3D localization performance compared 
to estimating depth from defocus and attains sub-millimeter resolution across all 
brightness levels tested.
The kaleidoscopic design's performance retains high accuracy and resolution 
even at lower brightness levels, demonstrating robustness to low photon counts and 
dark counts.

\noindent
\textbf{Limitations and future work.}
In this work, we only test one specific kaleidoscopic geometry with first-order 
reflections from one event.
Future work may test different kaleidoscopic geometries and higher-order reflections.
Developing a method to measure multiple events in one image would allow to 
construct a Compton camera or neutron scatter camera to localize the radiation 
source \cite{parajuli2022development,weinfurther2018model}.

Internal reflections at the scintillator's base surface could possibly 
appear in an image after reflecting off a mirrored surface.
An internal reflection has the same effect as adding a mirror to the 
scintillator's base surface. 
Then the following reflection off the physical mirror can be modeled as a 
second-order reflection.
Since internal reflection occurs at incidence angles higher than the critical 
angle, the mirror reflection of an internal reflection will exist away from 
first-order mirror reflections in $x$ and $y$ directions along the mirror's normal vector.
The camera's FOV in the experiments is limited so that imaging mirror reflections 
that follow from internal reflections at the scintillator's base surface is infrequent.
Hardware configurations with a larger camera FOV may capture internal reflections 
and have to account for them.
Imaging internal reflections will increase light collection and image complexity.

\section*{Acknowledgments}

This material is based upon work supported by the Department of Energy / National Nuclear Security Administration under Award Number(s) DE-NA0004196.

U.S. Department of Energy (Disclaimer): This work was prepared as an account of work sponsored by an agency of the United States Government. Neither the United States Government nor any agency thereof, nor any of their employees, nor any of their contractors, subcontractors or their employees, makes any warranty, express or implied, or assumes any legal liability or responsibility for the accuracy, completeness, or any third parties use or the results of such use of any information, apparatus, product, or process disclosed, or represents that its use would not infringe privately owned rights. Reference herein to any specific commercial product, process, or service by trade name, trademark, manufacturer, or otherwise, does not necessarily constitute or imply its endorsement, recommendation, or favoring by the United States Government or any agency thereof or its contractors or subcontractors. The views and opinions of authors expressed herein do not necessarily state or reflect those of the United States Government or any agency thereof, its contractors or subcontractors.

{
    \small
    \bibliographystyle{ieeenat_fullname}
    \bibliography{main}
}

\clearpage
\maketitlesupplementary

\appendix
\crefalias{section}{appendix}

\section{Theoretical Image Truncation Derivation} \label{sec:image_truncations_sup}

\begin{figure*}[!htbp]
\centering
\includegraphics[width=\linewidth]{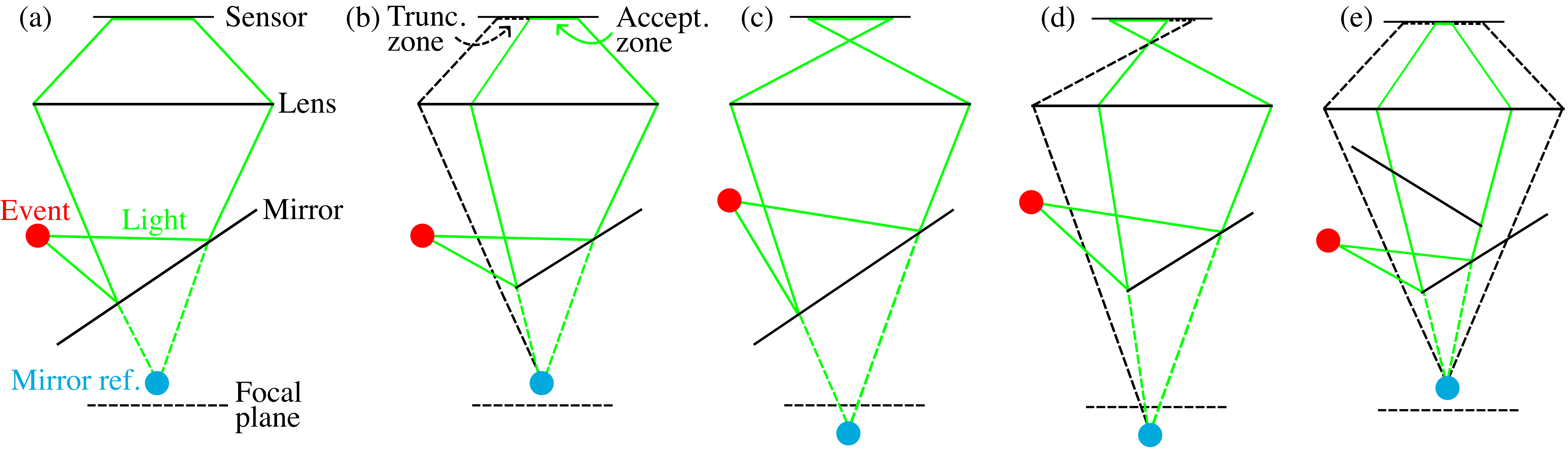}
\caption{\textbf{Mirror apertures and image truncations.} 
An event emits light onto a mirror that reflects into the camera. 
Some light from a mirror reflection cannot reach the sensor due to finite mirrors 
and defocus blur, forming a ``truncation zone" on the sensor.
The ``acceptance zone" denotes the area where light from a mirror reflection can 
arrive on the sensor.
Example cases include the following:
The mirror reflection is located within the focal plane (a,b,e) or beyond the 
focal plane (c,d).
All light that forms the mirror reflection reaches the sensor (a,c).
Some light that would form the mirror reflection is stopped at the mirror's edge 
and truncated on the sensor (b,d).
Some light that would form a second-order mirror reflection is stopped at both 
mirrors' edges and truncated on the sensor (e). 
The mirror for the second reflection in (e) is illustrated along the light's path.
Illustrations are drawn with scintillator index of refraction $n=1$.} 
\label{fig:trunc_theory}
\end{figure*}

An event's mirror reflection generated from mirror $k$ is located at $\bm{p_k}$ 
with an apparent location at $\bm{p_k}^{(a)}$.
Each edge of the mirror may impose a truncation line on the camera sensor.
A truncation line on the sensor, $l_\text{sensor}$, is determined as follows.
Denote a plane, $P_1$, that contains $\bm{p_k}$ and the mirror's edge.
If dealing with a second-order reflection or higher, the mirror's edge corresponds 
to the area where light can arrive on mirror $k$ after being stopped at a 
previous mirror, which may be different than mirror $k$'s physical edge.
Denote the line that intersects $P_1$ and the scintillator's base surface as $l_1$.
Compute the plane, $P_{\text{trunc}}$, that contains $l_1$ and $\bm{p_k}^{(a)}$.
Denote the side of $P_{\text{trunc}}$ that faces away from the mirror using the 
normal vector $\bm{n_{\text{trunc}}}$.
Compute the intersection between $P_{\text{trunc}}$ and the focal plane to be the
truncation line at the focal plane, $l_\text{focal}$.
Project $\bm{n_{\text{trunc}}}$ onto the focal plane, denoted as 
$\bm{n_{\text{focal}}}$.
Scale $l_\text{focal}$ by the magnification, $m=-\frac{S_2}{S_1}$, to obtain $l_\text{sensor}$.
Scale $\bm{n_{\text{focal}}}$ by $m$ to obtain $\bm{n_{\text{sensor}}}$.

The truncation side of $l_\text{sensor}$ where photons do not arrive depends on 
which side of the focal plane that $\bm{p_k}^{(a)}$ is located.
If $\bm{p_k}^{(a)}$ is beyond the focal plane away from the lens at a distance 
greater than $S_1$ from the lens, 
then the side of $l_\text{sensor}$ pointed to by $\bm{n_{\text{sensor}}}$ is 
truncated and contains no photon arrivals (\cref{fig:trunc_theory}b).
If $\bm{p_k}^{(a)}$ is between the focal plane and lens at a distance less than 
$S_1$ from the lens, 
then the side of $l_\text{sensor}$ opposite of $\bm{n_{\text{sensor}}}$ 
is truncated and contains no photon arrivals (\cref{fig:trunc_theory}d).

The area on the sensor where photons cannot arrive is the union of the
truncation sides of each truncation line.
We denote this area as the ``truncation zone" and its complement as the 
``acceptance zone" (\cref{fig:trunc_theory}b).
The event itself has no truncation zone on the sensor because its image is formed 
by light emitted directly to the camera without reflections.

\section{Imaging Theory Simulation Details} \label{sec:sim_details}
\cref*{fig:trunc_examples} in the main paper is generated with the following 
configuration using ray tracing and a thin lens.
The simulated scintillator has a 5.77 mm height, 20 mm base length, 120 degree 
opening angle, and index of refraction $n=1.5$.
A thin lens with 35 mm focal length and 25 mm diameter is placed at $S_1=45$ mm 
away from the focal plane at the scintillator's apex's apparent location.
A $512 \times 512$ sensor with 18.6 $\mu$m pixel pitch is placed at $S_2=157.5$ mm 
away from the lens.
100,000 photons are emitted isotropically from $(-0.5, 0.75, 1)$ mm (world 
coordinates).

\section{Gaussian Mixture Model Formulation} \label{sec:image_model_sup}
The complete-data likelihood function, $L$, for an image with $N$ photons, 
one event, and $K$ mirror reflections is
\begin{align}
L(\bm{\theta};\bm{t},\bm{z})=\prod_{i=1}^N \prod_{k=0}^K \left[ \pi_k \mathcal{N}(\bm{t_i};\bm{\mu_k},\sigma_k^2) \right]^{\mathbbm{1}{(z_i=k)}}
\end{align}
where $\bm{\theta}=(\bm{\mu},\bm{\sigma},\bm{\pi})$ are model parameters, 
and $\pi_k$ is the mixing weight for component $k$.
$\bm{t}=(\bm{t_1}, \bm{t_2}, ..., \bm{t_N})$ are the 2D coordinates of each of 
$N$ photon arrivals on the sensor and $\bm{z}=(z_1, z_2, ..., z_N)$ are the latent 
variables for which component a photon belongs to.
We apply a density-based weighting scheme to photon samples to minimize the 
influence of sparsely distributed dark counts.
The weighted complete-data likelihood function is 

\begin{align}
&L_w(\bm{\theta};\bm{t},\bm{z}) = \prod_{i=1}^N \prod_{k=0}^K \left[ \pi_k \mathcal{N}(\bm{t_i};\bm{\mu_k},\frac{1}{w_i}\sigma_k^2) \right]^{\mathbbm{1}{(z_i=k)}} \\
&= \prod_{i=1}^N \prod_{k=0}^K \left[ \pi_k \frac{w_i}{2\pi{\sigma_k}^2} \text{exp}\left( -\frac{w_i}{2{\sigma_k}^2} ||\bm{t_i}-\bm{\mu_k}||_2^2 \right) \right]^{\mathbbm{1}{(z_i=k)}}  \notag
\end{align}
where
\begin{align}
w_i = \sum_{j \in S_i^q} \text{exp} \left( -\frac{||\bm{t_i}-\bm{t_j}||_2^2}{\nu} \right)
\end{align}
is the weight assigned to photon $i$, $S_i^q$ is the set of $q$ 
nearest neighbors of photon $i$, and $\nu$ is a positive scalar.
The expected value of the weighted complete-data log-likelihood, $Q$, is
\begin{align} \label{eqn:Q_eqn}
Q & = E_{\bm{z}|\bm{t}}\left[\log L_w(\bm{\theta};\bm{t},\bm{z})\right] \\ & = \sum_i \sum_k r_{ik} \bigg[ \text{log}(\pi_k) + \text{log}(w_i) - \text{log}(2\pi{\sigma_k}^2) \notag \\ 
&\quad \quad \quad \quad \quad \quad - \frac{w_i}{2{\sigma_k}^2}||\bm{t_i}-\bm{\mu_k}||_2^2 \bigg] \notag
\end{align}

where
\begin{equation}
\begin{aligned}
r_{ik} & = E_{\bm{z}|\bm{t}}[\mathbbm{1}{(z_i=k)}] \\ & = \frac{\pi_k \mathcal{N}(\bm{t_i};\bm{\mu_k},{\sigma_k^2})}{\sum_{k'=0}^K \pi_{k'} \mathcal{N}(\bm{t_i};\bm{\mu_{k'}},{\sigma_{k'}^2})}
\end{aligned}
\end{equation}
gives the posterior distribution of $\bm{z}$.
$r_{ik}$ is the probability that photon $i$ comes from component $k$, given 
the current parameter values.

\section{Experimental Parameter Values} \label{sec:param_values}
In all experiments, we use the regularization term with $\lambda=10$.
We use $q=10$ nearest neighbors and $\nu=10$ pixels for assigning photon weights.
Up to ten possible event locations in the scintillator are evaluated in the 
initialization procedure equispaced over the scintillator's depth.
During one M-step, we run gradient ascent for 1,000 steps with a step size of 1e-7.
We run the EM algorithm until the distance in the estimated event location between 
consecutive steps is less than 0.01 mm, or until 100 steps are taken.
We clip all $\sigma_k$'s to a a minimum value of 10 pixels to account for 
imperfect focusing and a non-ideal point source.
The gradient of $\sigma_k$ is set to 0 if $\sigma_k$ gets clipped to the minimum value.
We set $A=41.7$ mm, $S_1=45$ mm, $S_2=72$ mm, and $a=0.25$.
See \cref{sec:param_cal} for how we calibrate these camera parameters.
The focal plane is set to the scintillator's apparent depth at $z_w=2.75$ mm.

\section{Experimental Camera Parameter Calibration} \label{sec:param_cal}
\begin{figure}[!htbp]
\centering
\includegraphics[width=.5\linewidth]{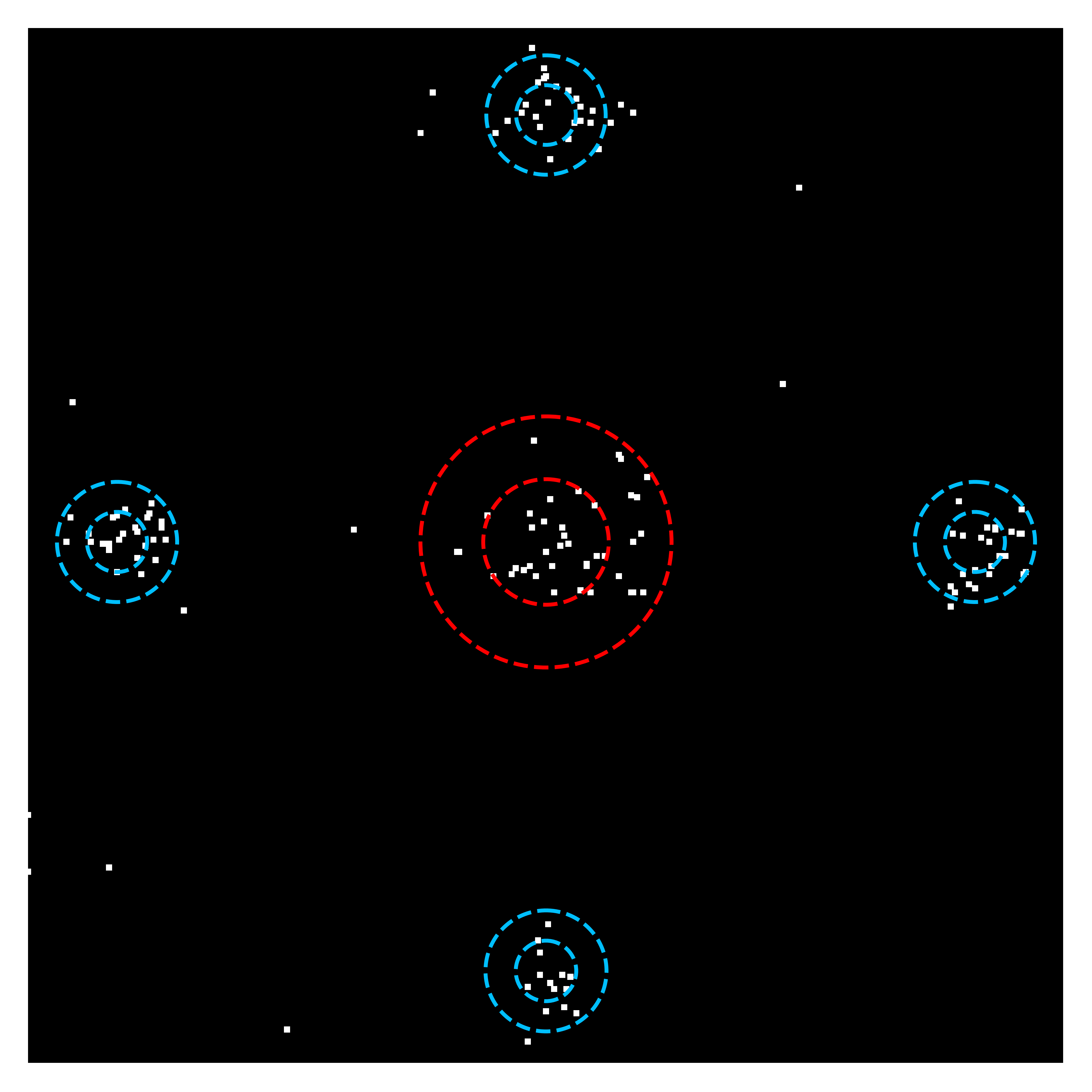}
\caption{\textbf{Selected experimental calibration image.} The image is 
overlaid with the Gaussian components found after manually adjusting the 
camera parameters.
Each dashed red circle is centered on the Gaussian component's mean. 
The inner and outer circles are one and two standard deviations in radius, respectively.
Red and blue circles represent the event and mirror reflections, respectively.
Pixels with a photon are enlarged with a $3 \times 3$ filter for visualization purposes.
} 
\label{fig:calibration}
\end{figure}

We select the experimental image shown in \cref{fig:calibration} 
where the event is approximately centered and mirror reflections are near the 
image's edges.
The camera's field of view covers about 5.5 mm at the focal plane, and the +$x$ 
mirror reflection's photon cluster is approximately centered at pixel coordinate $(470,256)$.
We approximate the observed mirror reflection's $x$-coordinate in world 
coordinates as $470/512 \times 5.5/2=2.52$ mm. 
We assign the event's location to be $(0,0,2.9)$ mm.
This event location and the scintillator's geometry produce a +$x$ mirror 
reflection located at $x=2.52$ mm, matching the observed mirror reflection.
The event location is fixed, and the camera aperture diameter is set to $A=50/1.2=41.7$ mm.
The camera parameter values for $S_1$, $S_2$, and $a$ are manually adjusted until 
the resulting Gaussian components ($\bm{\mu_k}$,$\sigma_k$) coincide with the 
event and mirror reflection photon clusters.  
Values are set so that approximately all photons in the photon cluster are 
contained within a radius of 2$\sigma_k$ from component $k$'s centroid (plotted 
in \cref{fig:calibration}).
Dark counts are ignored.
The resulting values are $S_1=45$ mm, $S_2=72$ mm, and $a=0.25$.
The focal plane is set to the scintillator's apparent depth at 
$z_w=5.77-5.77/1.91=2.75$ mm.
The lens plane is set at $z_w=2.75+45=47.75$ mm. 
The sensor plane is set at $z_w=47.75+72=119.75$ mm.

\section{Ablation Study: Regularization} \label{sec:ablation}
\begin{figure*}
\centering
\includegraphics[width=\linewidth]{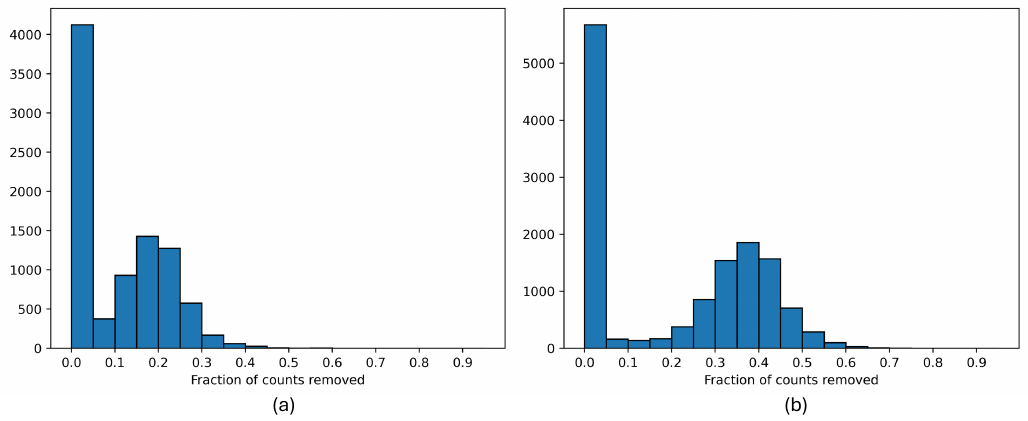}
\caption{\textbf{Ablation study. Experimental fraction of counts in an image removed during mirror removals without regularization ($\lambda=0$).} (a) One mirror reflection removal. Median, mean, stdev: 0.10, 0.10, and 0.11. 8,972 images.  
(b) Two mirror reflection removals. Median, mean, stdev: 0.27, 0.21, and 0.19. 13,458 images.
} 
\label{fig:removal_frac_no_reg}
\end{figure*}

\begin{figure}
\centering
\includegraphics[width=\linewidth]{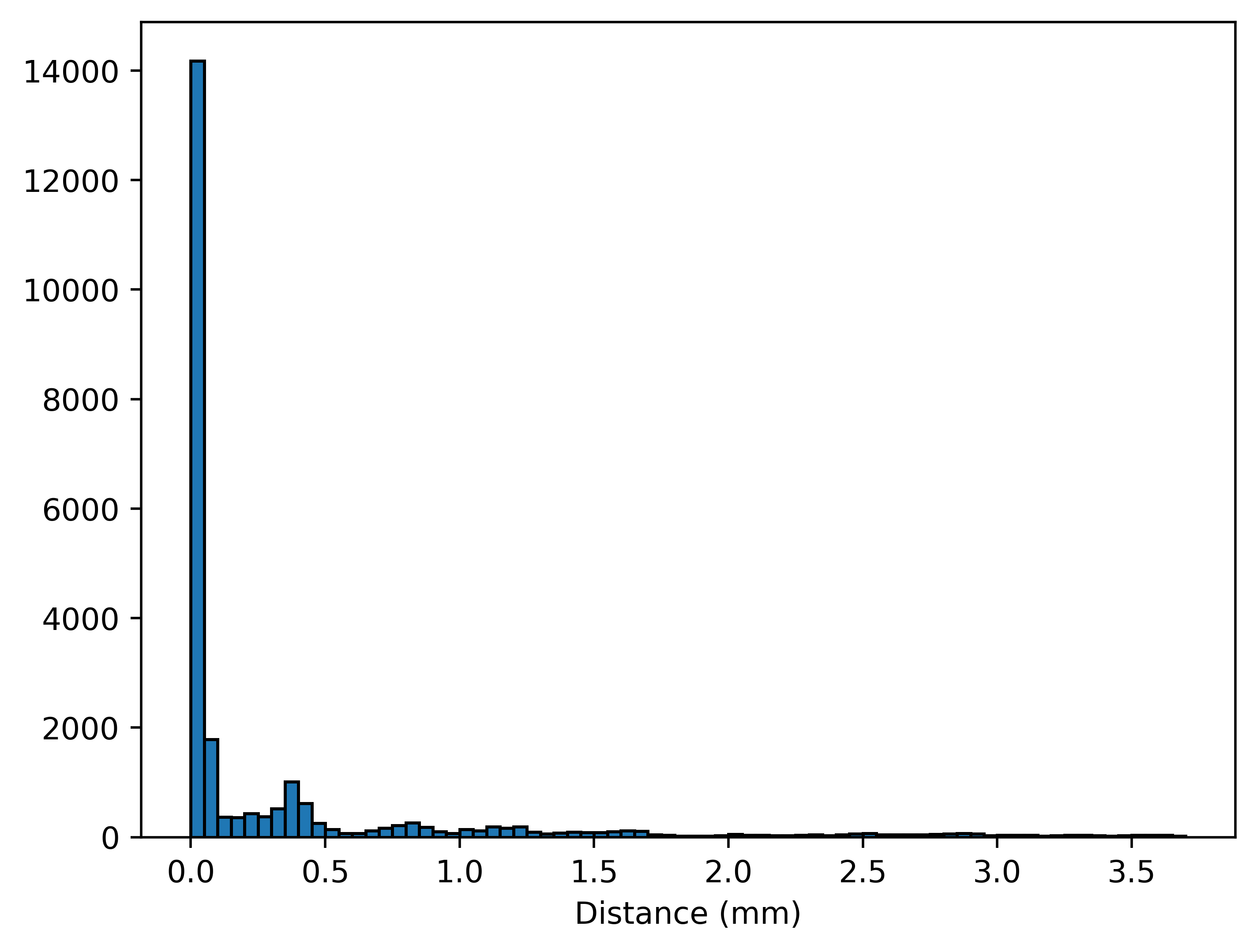}
\caption{\textbf{Ablation study. Experimental agreement in event location measurements without regularization ($\lambda=0$).} Histogram of distances 
between mean estimated event location and each image's estimated event location 
after mirror reflection removals. Median, mean, stdev: 0.03 mm, 0.39 mm, and 0.79 mm. 24,673 distances.
} 
\label{fig:crossval_error_no_reg}
\end{figure}

\begin{figure}[!htbp]
\centering
\includegraphics[width=\linewidth]{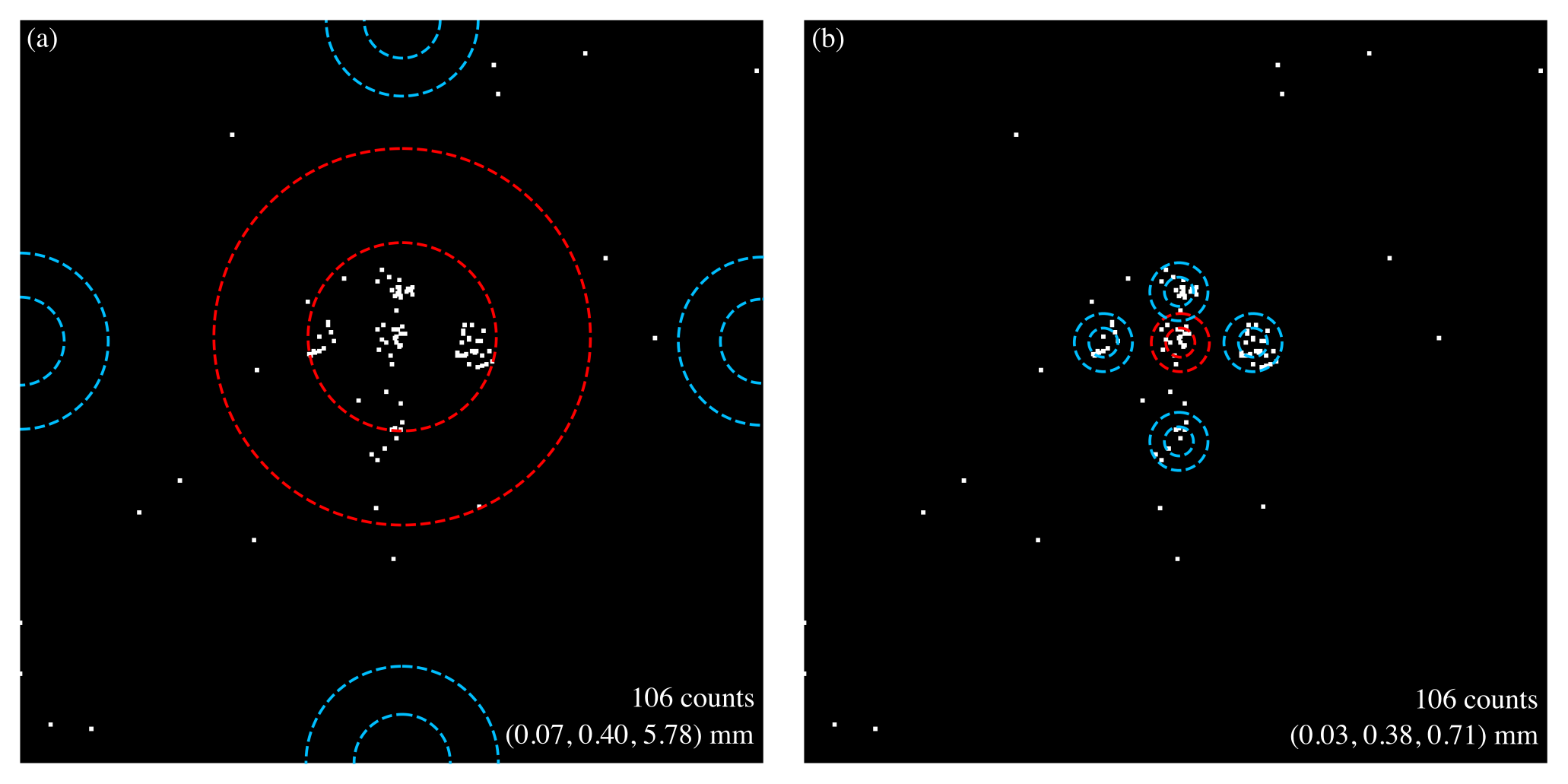}
\caption{\textbf{Regularization example.} An experimental image overlaid with the 
algorithm's estimated Gaussian components with (a) $\lambda=0$ (no regularization) and (b) $\lambda=10$ (regularization).
Each dashed red circle is centered on the Gaussian component's mean. 
The inner and outer circles are one and two standard deviations in radius, respectively.
Red and blue circles represent the event and mirror reflections, respectively.
The number of counts in the image and the algorithm's estimated event location are
shown in each image.
Pixels with a photon are enlarged with a $3 \times 3$ filter for visualization purposes.
} 
\label{fig:regularization}
\end{figure}

We perform the same experiments as in \cref*{sec:exp} of the main paper on the 
same 4,379 experimental images containing at least 60 counts, except we set $\lambda=0$ (no regularization).
The results are reported in \cref{fig:crossval_error_no_reg,fig:removal_frac_no_reg}. 
An elevated frequency of images where no photons are removed during mirror 
reflection removals is observed in \cref{fig:removal_frac_no_reg}.
For these images, the algorithm is likely converging to an event location with an 
erroneously high $z_w$ coordinate. 
This results in a Gaussian component with a large $\sigma_0$ that covers photon 
clusters of multiple mirror reflections, as shown in \cref{fig:regularization}a.
The high frequency of short distances in \cref{fig:crossval_error_no_reg} also 
indicates that mirror reflections are not being properly identified and removed, 
and the algorithm is converging to the same estimate over multiple identical images.
Thus, we observe a decrease in localization performance without regularization.

\section{Denoising and Estimating Depth from Defocus (Prior Method)} \label{sec:prior_method}
Prior methods that estimate a 3D event location using depth from defocus have also
attempted to remove dark counts using a denoising algorithm before solving for the 
location's maximum likelihood estimate \cite{bocchieri2024scintillation}.
The denoising algorithm consists of computing the minimum spanning tree among 
all photons in an image and removing edges longer than a chosen threshold, $T_{edge}$. 
The photons in the largest connected component by cardinality are classified as 
the scintillation photons, and the rest are discarded as dark counts. 
We test this method on the simulated non-kaleidoscopic images 
(described in \cref*{sec:simulations} of the main paper) and report results in \cref{tab:sim_results_supp}. 
We observe similar performance as the non-kaleidoscopic version of our proposed algorithm.

\begin{table}[!htbp]
\footnotesize
\centering
\begin{tabular}{cccccc} 
 \toprule
 & & \multicolumn{4}{c}{Average error and resolution (mm) $\downarrow$} \\
 \cmidrule(lr){3-6}
 $N_0$ & $T_{edge}$ (pixels) & 3D Error & x Res. & y Res. & z Res. \\
 \midrule
  \multirow{3}{*}{30} & 40 & 0.81 & 1.43  & 1.41 & 1.42  \\
  & 80  & 0.64 & 1.04 & 1.02 & 1.04   \\
  & 120  & 1.26 & 1.72 & 1.91 & 1.73   \\
  \midrule
  \multirow{3}{*}{20} & 40 & 0.92  & 1.59 & 1.56 & 1.56  \\
  & 80  & 0.68 & 1.05 & 1.03 & 1.06   \\
  & 120  & 1.59 & 2.10 & 2.38 & 2.21   \\
  \midrule
  \multirow{3}{*}{10} & 40 & 1.14 & 1.89  & 1.85 & 1.90 \\
  & 80  & 0.80 & 1.20 & 1.13 & 1.14  \\
  & 120  & 2.04 & 2.84 & 2.90 & 2.83   \\

 \bottomrule
\end{tabular}
\caption{\textbf{Prior method all simulation results.} 
Estimating depth from defocus on non-kaleidoscopic images after attempting to 
remove dark counts as in a prior method \cite{bocchieri2024scintillation}.}
\label{tab:sim_results_supp}
\end{table}

\clearpage
\onecolumn

\section{Additional Figures}

\begin{figure}[!htbp]
\centering
\includegraphics[width=.5\linewidth]{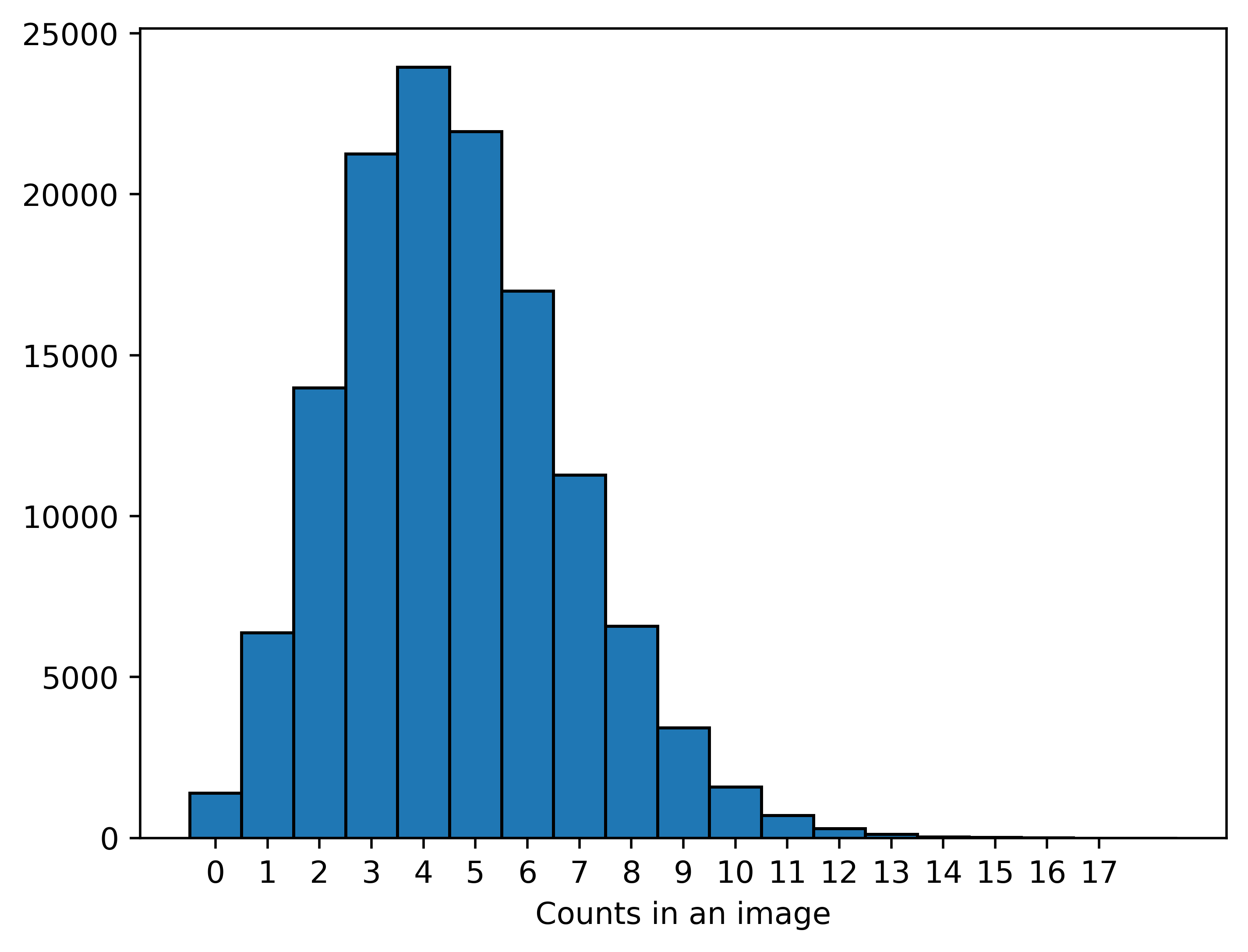}
\caption{\textbf{Histogram of dark counts per experimental image.} 
A median of 4 dark counts per image after zeroing hot pixels is observed out of 
130,000 images taken in the dark with no gamma-ray source present.} 
\label{fig:dark_counts_hist}
\end{figure}

\begin{figure}[!htbp]
\centering
\includegraphics[width=\linewidth]{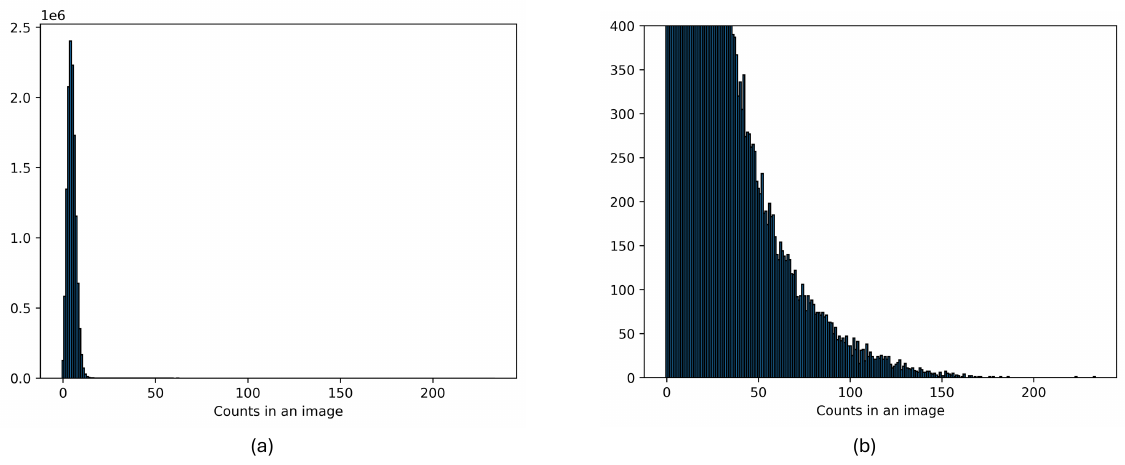}
\caption{\textbf{Histogram of counts per experimental image with the gamma-ray source present.} A median of 4 counts 
per image after zeroing hot pixels is observed out of 13,000,000 images taken with 
the gamma-ray source present. 
a) The full histogram. 
b) The histogram clipped in the y-axis.} 
\label{fig:cap_counts_hist}
\end{figure}

\begin{figure}[!htbp]
\centering
\includegraphics[width=\linewidth]{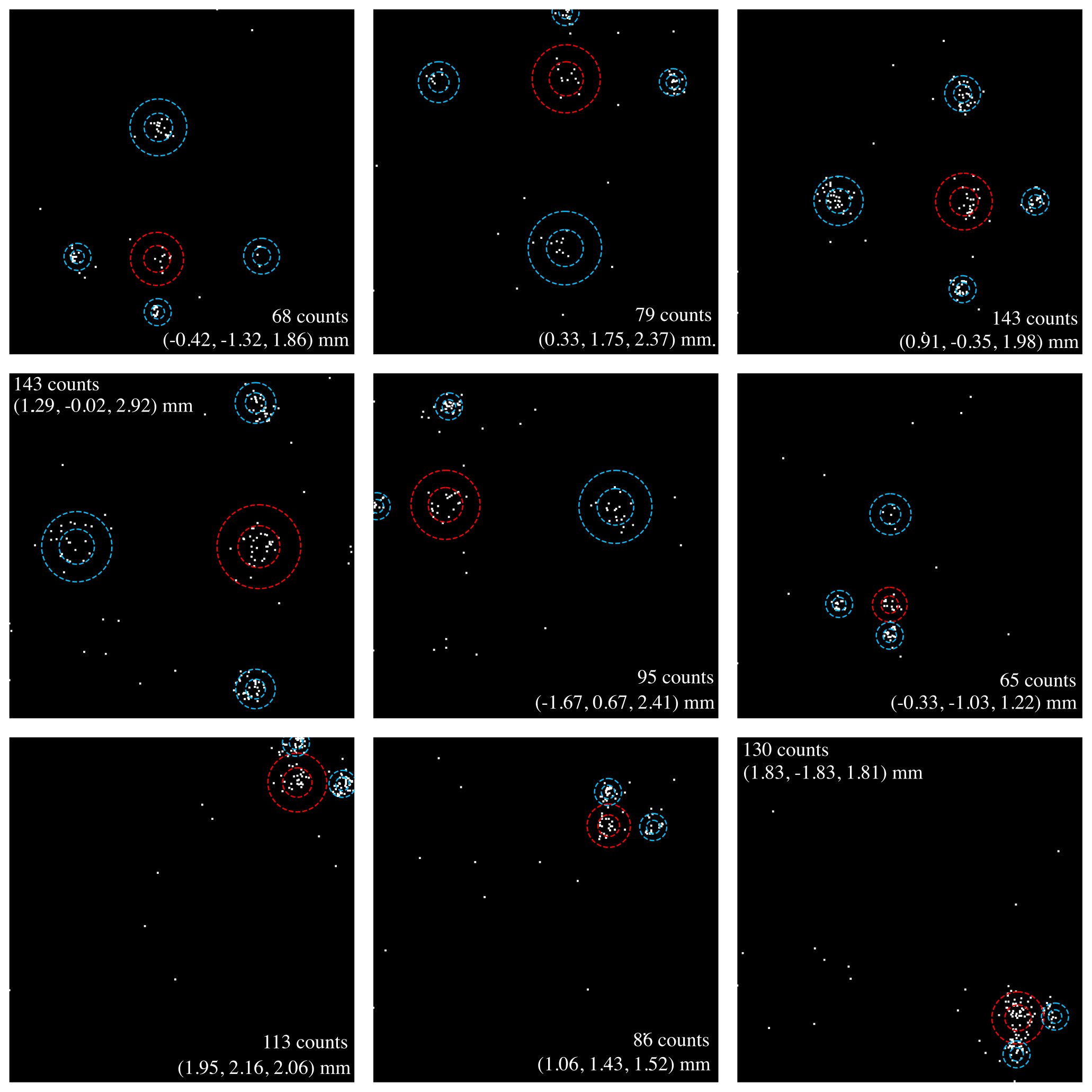}
\caption{\textbf{Additional selected experimental images.} Experimental images overlaid with the algorithm's estimated Gaussian components.
Each dashed red circle is centered on the Gaussian component's mean. 
The inner and outer circles are one and two standard deviations in radius, respectively.
Red and blue circles represent the event and mirror reflections, respectively.
Pixels with a photon are enlarged with a $3 \times 3$ filter for visualization purposes.
The number of counts in the image and the algorithm's estimated event location are
shown in each image.
} 
\label{fig:example_figures_sup}
\end{figure}

\begin{figure}[!htbp]
\centering
\includegraphics[width=.5\linewidth]{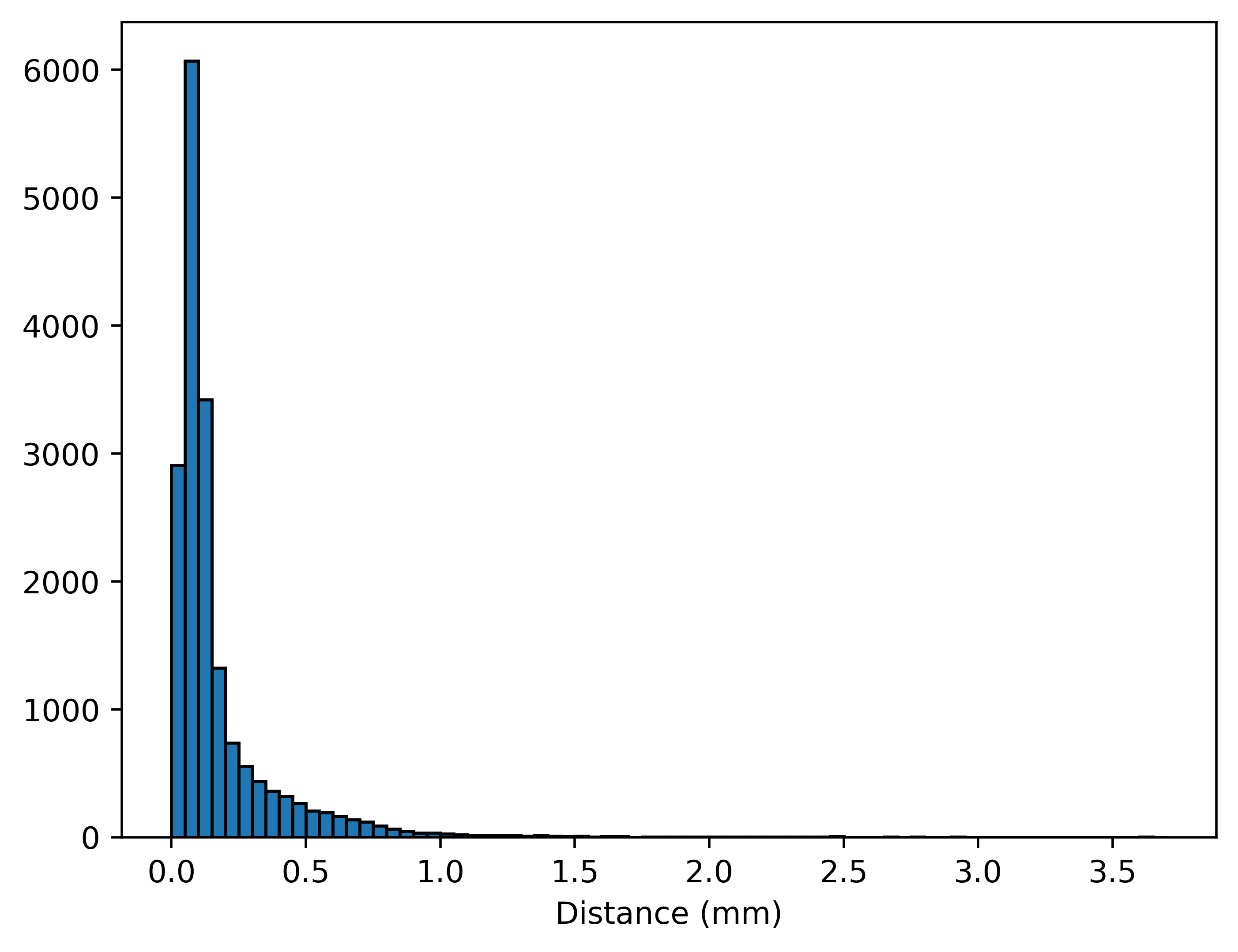}
\caption{\textbf{Experimental agreement in event location measurements using regularization ($\lambda=10$).} Histogram of distances 
between mean estimated event location and each image's estimated event location 
after mirror reflection removals. Median, mean, stdev: 0.10 mm, 0.17 mm, and 0.22 mm. 17,666 distances.
} 
\label{fig:crossval_error}
\end{figure}

\begin{figure}[!htbp]
\centering
\includegraphics[width=\linewidth]{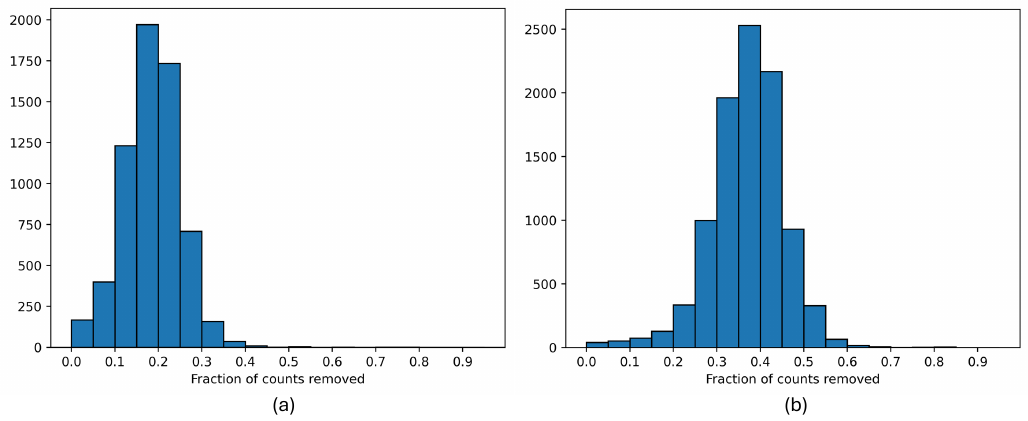}
\caption{\textbf{Experimental fraction of counts in an image removed during mirror removals using regularization ($\lambda=10$).} (a) One mirror reflection removal. Median, mean, stdev: 0.19, 0.18, and 0.07. 6,424 images.  
(b) Two mirror reflection removals. Median, mean, stdev: 0.38, 0.37, and 0.08. 9,636 images.
} 
\label{fig:removal_frac}
\end{figure}

\begin{figure}[!htbp]
\centering
\includegraphics[width=.5\linewidth]{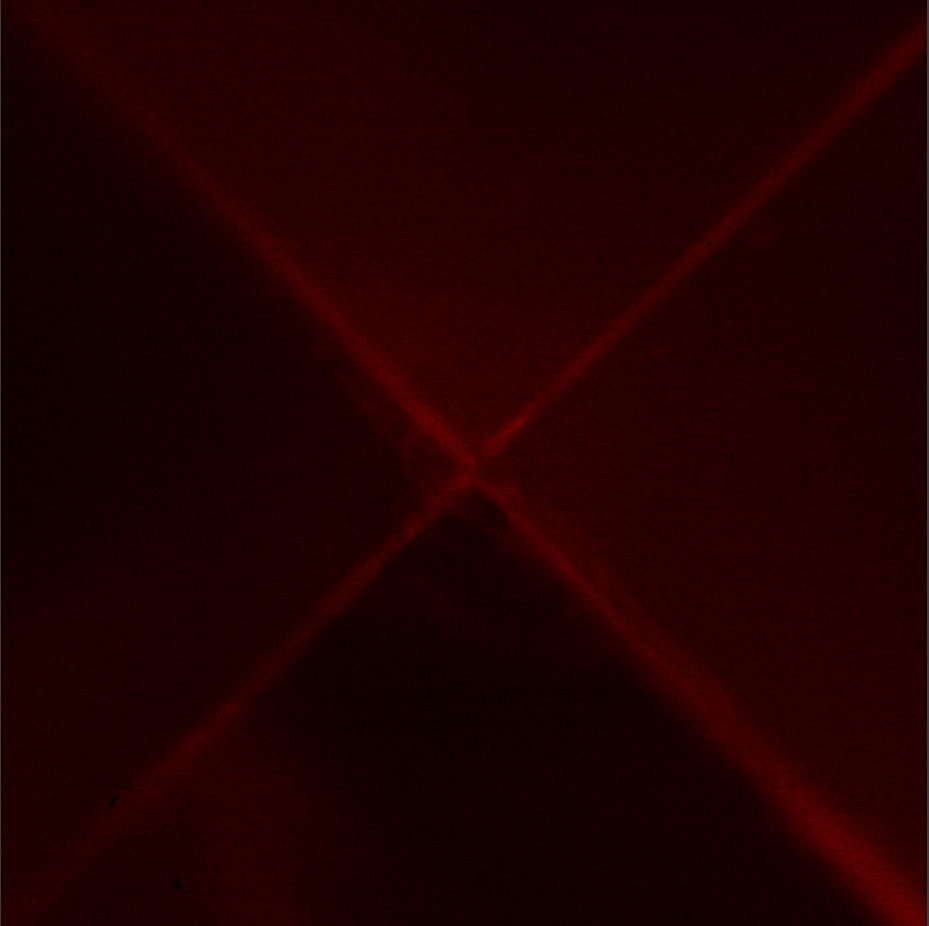}
\caption{\textbf{Experimental focus.} A view of how the camera is focused on the scintillator in the experiments. The edges between mirrors are visible.} 
\label{fig:experiment_focus}
\end{figure}

\clearpage
\section{Gradient Derivations} \label{sec:grad}

Consider a square pyramid kaleidoscopic scintillator with height $h$ and index of refraction $n$. 
The camera's focal plane is located at the apparent depth of the scintillator's apex at $z_w = h - h/n$ (world coordinates).
Denote an event's real location $\bm{p_0}$ at $(x_0,y_0,z_{w0})$ in world coordinates.
To transform from world coordinates (origin at pyramid apex) to camera 
coordinates (origin at lens), apply $z_c = z_{lens} - z_w$, where $z_{lens}$ is in world coordinates.
The event's mirror reflection $k$ is located at $\bm{p_k}=(x_k,y_k,z_k)=\bm{T_k}\bm{p_0}$.
$\alpha_{xyk}$ denotes the coefficient in $\bm{T_k}$ corresponding to $x_0$ and $y_k$.
$k=0$ refers to the event itself, and not a mirror reflection.
These derivations assume $\bm{\mu_k} = \left[\frac{S_2}{z_k} x_k, \frac{S_2}{z_k} y_k\right]$. 
In these derivations, we use camera coordinates (origin at lens) denoted as $z_k$ instead of $z_{ck}$ as in the main paper. 
The camera sees events and mirror reflections at their apparent depth rather than their real depth.  
$z_{w0}$ is the $z$-coordinate of the event's real location in world coordinates, and $z_k$ is the event or mirror reflection's apparent $z$-coordinate in camera coordinates.

\subsection{Unweighted Likelihood} \label{sec:like}

\begin{align}
x_k = \alpha_{xxk}x_0 + \alpha_{xyk}y_0 + \alpha_{xzk}z_{w0}
\end{align}
\begin{align}
y_k = \alpha_{yxk}x_0 + \alpha_{yyk}y_0 + \alpha_{yzk}z_{w0}
\end{align}
\begin{align}
z_k = z_\text{lens} - \left(h - \frac{h - (\alpha_{zxk}x_0 + \alpha_{zyk}y_0 + \alpha_{zzk}z_{w0})}{n}\right)
\end{align}

\begin{align}
\bm{\mu_k} = \left[\frac{S_2}{z_k} x_k, \frac{S_2}{z_k} y_k\right]
\end{align}
\begin{align}
\sigma_k = a \frac{AS_2}{S_1} \frac{|S_1-z_k|}{z_k}
\end{align}

\begin{align}
L(\bm{\theta};\bm{t},\bm{z}) &= \prod_{i=1}^N \prod_{k=0}^K \left[ \pi_k \mathcal{N}(\bm{t_i};\bm{\mu_k},\sigma_k^2) \right]^{\mathbbm{1}{(z_i=k)}} \\
&= \prod_{i=1}^N \prod_{k=0}^K \left[ \pi_k \frac{1}{2\pi{\sigma_k}^2} \text{exp}\left( -\frac{1}{2{\sigma_k}^2} (\bm{t_i}-\bm{\mu_k})^T(\bm{t_i}-\bm{\mu_k}) \right) \right]^{\mathbbm{1}{(z_i=k)}}
\end{align}

\begin{align}
(\bm{t_i}-\bm{\mu_k})^T(\bm{t_i}-\bm{\mu_k}) &= {t_{ix}}^2 - 2t_{ix}\left(S_2\frac{x_k}{z_{k}}\right) + \left(S_2\frac{x_k}{z_{k}}\right)^2 + {t_{iy}}^2 - 2t_{iy}\left(S_2\frac{y_k}{z_{k}}\right) + \left(S_2\frac{y_k}{z_{k}}\right)^2 \\
&= {t_{ix}}^2 - 2t_{ix}S_2\frac{x_k}{z_{k}} + {S_2}^2\left(\frac{x_k}{z_{k}}\right)^2 + {t_{iy}}^2 - 2t_{iy}S_2\frac{y_k}{z_{k}} + {S_2}^2\left(\frac{y_k}{z_{k}}\right)^2
\end{align}

\begin{align}
Q = \sum_i \sum_k r_{ik} \left[ \text{log}\pi_k - \text{log}(2\pi{\sigma_k}^2) - \frac{{\sigma_k}^{-2}}{2}(\bm{t_i}-\bm{\mu_k})^T(\bm{t_i}-\bm{\mu_k}) \right]
\end{align}

\begin{align}
\frac{x_k}{z_k} = \frac{\alpha_{xxk}x_0 + \alpha_{xyk}y_0 + \alpha_{xzk}z_{w0}}{z_\text{lens} - \left(h - \frac{h - (\alpha_{zxk}x_0 + \alpha_{zyk}y_0 + \alpha_{zzk}z_{w0})}{n}\right)}
\end{align}

\begin{align}
\frac{y_k}{z_k} = \frac{\alpha_{yxk}x_0 + \alpha_{yyk}y_0 + \alpha_{yzk}z_{w0}}{z_\text{lens} - \left(h - \frac{h - (\alpha_{zxk}x_0 + \alpha_{zyk}y_0 + \alpha_{zzk}z_{w0})}{n}\right)}
\end{align}

$u_x = x_k = \alpha_{xxk}x_0 + \alpha_{xyk}y_0 + \alpha_{xzk}z_{w0}$

${u_x}'(x_0) = \alpha_{xxk}$

${u_x}'(y_0) = \alpha_{xyk}$

${u_x}'(z_{w0}) = \alpha_{xzk}$

$u_y = y_k = \alpha_{yxk}x_0 + \alpha_{yyk}y_0 + \alpha_{yzk}z_{w0}$

${u_y}'(x_0) = \alpha_{yxk}$

${u_y}'(y_0) = \alpha_{yyk}$

${u_y}'(z_{w0}) = \alpha_{yzk}$

$v_z = z_k = z_\text{lens} - \left(h - \frac{h - (\alpha_{zxk}x_0 + \alpha_{zyk}y_0 + \alpha_{zzk}z_{w0})}{n}\right)$

${v_z}'(x_0) = -\alpha_{zxk}/n$

${v_z}'(y_0) = -\alpha_{zyk}/n$

${v_z}'(z_{w0}) = -\alpha_{zzk}/n$

\begin{align}
\frac{\partial}{\partial x_0}\left(\frac{x_k}{z_k}\right) = \frac{{u_x}'v_z -u_x{v_z}'}{v^2}  = \frac{\alpha_{xxk} z_k + \alpha_{zxk} x_k/n}{{z_k}^2}
\end{align}

\begin{align}
\frac{\partial}{\partial y_0}\left(\frac{x_k}{z_k}\right) = \frac{\alpha_{xyk} z_k + \alpha_{zyk} x_k/n}{{z_k}^2}
\end{align}

\begin{align}
\frac{\partial}{\partial z_{w0}}\left(\frac{x_k}{z_k}\right) = \frac{\alpha_{xzk} z_k + \alpha_{zzk} x_k/n}{{z_k}^2}
\end{align}

\begin{align}
\frac{\partial}{\partial x_0}\left(\frac{y_k}{z_k}\right) = \frac{{u_y}'v_z -u_y{v_z}'}{v^2}  = \frac{\alpha_{yxk} z_k + \alpha_{zxk} y_k/n}{{z_k}^2}
\end{align}

\begin{align}
\frac{\partial}{\partial y_0}\left(\frac{y_k}{z_k}\right) = \frac{\alpha_{yyk} z_k + \alpha_{zyk} y_k/n}{{z_k}^2}
\end{align}

\begin{align}
\frac{\partial}{\partial z_{w0}}\left(\frac{y_k}{z_k}\right) = \frac{\alpha_{yzk} z_k + \alpha_{zzk} y_k/n}{{z_k}^2}
\end{align}

\begin{align}
\frac{\partial}{\partial x_0}\left(\left(\frac{x_k}{z_k}\right)^2\right) = 2 \frac{x_k}{z_k} \frac{\partial}{\partial x_0}\left(\frac{x_k}{z_k}\right)
\end{align}

\begin{align}
\frac{\partial}{\partial y_0}\left(\left(\frac{x_k}{z_k}\right)^2\right) = 2 \frac{x_k}{z_k} \frac{\partial}{\partial y_0}\left(\frac{x_k}{z_k}\right)
\end{align}

\begin{align}
\frac{\partial}{\partial z_{w0}}\left(\left(\frac{x_k}{z_k}\right)^2\right) = 2 \frac{x_k}{z_k} \frac{\partial}{\partial z_{w0}}\left(\frac{x_k}{z_k}\right)
\end{align}

\begin{align}
\frac{\partial}{\partial x_0}\left(\left(\frac{y_k}{z_k}\right)^2\right) = 2 \frac{y_k}{z_k} \frac{\partial}{\partial x_0}\left(\frac{y_k}{z_k}\right)
\end{align}

\begin{align}
\frac{\partial}{\partial y_0}\left(\left(\frac{y_k}{z_k}\right)^2\right) = 2 \frac{y_k}{z_k} \frac{\partial}{\partial y_0}\left(\frac{y_k}{z_k}\right)
\end{align}

\begin{align}
\frac{\partial}{\partial z_{w0}}\left(\left(\frac{y_k}{z_k}\right)^2\right) = 2 \frac{y_k}{z_k} \frac{\partial}{\partial z_{w0}}\left(\frac{y_k}{z_k}\right)
\end{align}

\begin{align}
\sigma_k = a \frac{AS_2}{S_1} \frac{|S_1-z_k|}{z_k}
\end{align}

$ \frac{|S_1-z_k|}{z_k} = \frac{|S_1 - z_\text{lens} - \left(h - \frac{h - (\alpha_{zxk}x_0 + \alpha_{zyk}y_0 + \alpha_{zzk}z_{w0})}{n}\right)|}{z_\text{lens} - \left(h - \frac{h - (\alpha_{zxk}x_0 + \alpha_{zyk}y_0 + \alpha_{zzk}z_{w0})}{n}\right)} $

$ u = S_1 - z_k = S_1 - z_\text{lens} - \left(h - \frac{h - (\alpha_{zxk}x_0 + \alpha_{zyk}y_0 + \alpha_{zzk}z_{w0})}{n}\right)$

\begin{align}
\frac{\partial \sigma_k}{\partial x_0} = a\frac{AS_2}{S_1} \frac{\partial}{\partial x_0} \frac{|S_1-z_k|}{z_k} =  a\frac{AS_2}{S_1} \frac{\alpha_{zxk}sign(u)z_k/n - |u|(-\alpha_{zxk}/n)}{{z_k}^2} =  a\frac{AS_2}{S_1} \frac{\alpha_{zxk}sign(u)z_k/n + \alpha_{zxk}|u|/n}{{z_k}^2}
\end{align}

\begin{align}
\frac{\partial \sigma_k}{\partial y_0} = a\frac{AS_2}{S_1} \frac{\partial}{\partial y_0} \frac{|S_1-z_k|}{z_k} =  a\frac{AS_2}{S_1} \frac{\alpha_{zyk}sign(u)z_k/n + \alpha_{zyk}|u|/n}{{z_k}^2}
\end{align}

\begin{align}
\frac{\partial \sigma_k}{\partial z_{w0}} =  a\frac{AS_2}{S_1} \frac{\partial}{\partial z_{w0}} \frac{|S_1-z_k|}{z_k} =   a\frac{AS_2}{S_1} \frac{\alpha_{zzk}sign(u)z_k/n + \alpha_{zzk}|u|/n}{{z_k}^2}
\end{align}

\begin{align} \label{eqn:Q_no_weight}
Q &= \sum_i \sum_k r_{ik} \left[ \text{log}\pi_k - \text{log}(2\pi{\sigma_k}^2) - \frac{{\sigma_k}^{-2}}{2}(\bm{t_i}-\bm{\mu_k})^T(\bm{t_i}-\bm{\mu_k}) \right] \\
&= \sum_i \sum_k r_{ik} \left[ \text{log}\pi_k - \text{log}(2\pi{\sigma_k}^2) - \frac{{\sigma_k}^{-2}}{2} \left({t_{ix}}^2 - 2t_{ix}S_2\frac{x_k}{z_k} + {S_2}^2\left(\frac{x_k}{z_k}\right)^2 + {t_{iy}}^2 - 2t_{iy}S_2\frac{y_k}{z_k} + {S_2}^2\left(\frac{y_k}{z_k}\right)^2 \right) \right]
\end{align}

\begin{equation}
f(x_k,y_k,z_k) = \left({t_{ix}}^2 - 2t_{ix}S_2\frac{x_k}{z_k} + {S_2}^2\left(\frac{x_k}{z_k}\right)^2 + {t_{iy}}^2 - 2t_{iy}S_2\frac{y_k}{z_k} + {S_2}^2\left(\frac{y_k}{z_k}\right)^2 \right)
\end{equation}

\begin{equation}
\frac{\partial f}{\partial x_0} = -2t_{ix}S_2 \frac{\partial}{\partial x_0} \left(\frac{x_k}{z_k}\right) + {S_2}^2 \frac{\partial}{\partial x_0}\left(\left(\frac{x_k}{z_k}\right)^2\right) - 2t_{iy}S_2 \frac{\partial}{\partial x_0} \left(\frac{y_k}{z_k}\right) + {S_2}^2 \frac{\partial}{\partial x_0}\left(\left(\frac{y_k}{z_k}\right)^2\right)
\end{equation}

\begin{equation}
\frac{\partial f}{\partial y_0} = -2t_{ix}S_2 \frac{\partial}{\partial y_0} \left(\frac{x_k}{z_k}\right) + {S_2}^2 \frac{\partial}{\partial y_0}\left(\left(\frac{x_k}{z_k}\right)^2\right) - 2t_{iy}S_2 \frac{\partial}{\partial y_0} \left(\frac{y_k}{z_k}\right) + {S_2}^2 \frac{\partial}{\partial y_0}\left(\left(\frac{y_k}{z_k}\right)^2\right)
\end{equation}

\begin{equation}
\frac{\partial f}{\partial z_{w0}} = -2t_{ix}S_2 \frac{\partial}{\partial z_{w0}} \left(\frac{x_k}{z_k}\right) + {S_2}^2 \frac{\partial}{\partial z_{w0}}\left(\left(\frac{x_k}{z_k}\right)^2\right) - 2t_{iy}S_2 \frac{\partial}{\partial z_{w0}} \left(\frac{y_k}{z_k}\right) + {S_2}^2 \frac{\partial}{\partial z_{w0}}\left(\left(\frac{y_k}{z_k}\right)^2\right)
\end{equation}

\begin{align} \label{eqn:dx_no_weight}
\frac{\partial Q}{\partial x_0} = \sum_i \sum_k r_{ik} \left[ -2{\sigma_k}^{-1}\frac{\partial \sigma_k}{\partial x_0} - \left( -{\sigma_k}^{-3} \frac{\partial \sigma_k}{\partial x_0}f + \frac{{\sigma_k}^{-2}}{2} \frac{\partial f}{\partial x_0}  \right) \right] 
\end{align}

\begin{align} \label{eqn:dy_no_weight}
\frac{\partial Q}{\partial y_0} = \sum_i \sum_k r_{ik} \left[ -2{\sigma_k}^{-1}\frac{\partial \sigma_k}{\partial y_0} - \left( -{\sigma_k}^{-3} \frac{\partial \sigma_k}{\partial y_0}f + \frac{{\sigma_k}^{-2}}{2} \frac{\partial f}{\partial y_0}  \right) \right] 
\end{align}

\begin{align} \label{eqn:dz_no_weight}
\frac{\partial Q}{\partial z_{w0}} = \sum_i \sum_k r_{ik} \left[ -2{\sigma_k}^{-1}\frac{\partial \sigma_k}{\partial z_{w0}} - \left( -{\sigma_k}^{-3} \frac{\partial \sigma_k}{\partial z_{w0}}f + \frac{{\sigma_k}^{-2}}{2} \frac{\partial f}{\partial z_{w0}}  \right) \right] 
\end{align}

\subsection{Weighted Likelihood} \label{sec:weighted_like}

$w_i$ is the weight assigned to photon sample $i$.
$w_i = \sum\limits_{j \in S_i^q} \text{exp} \left( -\frac{||\bm{t_i}-\bm{t_j}||_2^2}{\nu} \right)$
where $S_i^q$ is the set of $q$ nearest neighbors of photon $i$, and $\nu$ is a 
positive scalar.

\begin{align}
L(\bm{\theta};\bm{t},\bm{z}) &= \prod_{i=1}^N \prod_{k=0}^K \left[ \pi_k \mathcal{N}(\bm{t_i};\bm{\mu_k},\frac{1}{w_i}\sigma_k^2) \right]^{\mathbbm{1}{(z_i=k)}} \\
&= \prod_{i=1}^N \prod_{k=0}^K \left[ \pi_k \frac{w_i}{2\pi{\sigma_k}^2} \text{exp}\left( -\frac{w_i}{2{\sigma_k}^2} (\bm{t_i}-\bm{\mu_k})^T(\bm{t_i}-\bm{\mu_k}) \right) \right]^{\mathbbm{1}{(z_i=k)}}
\end{align}

\begin{align}
Q = \sum_i \sum_k r_{ik} \left[ \text{log}(\pi_k) + \text{log}(w_i) - \text{log}(2\pi{\sigma_k}^2) - \frac{w_i{\sigma_k}^{-2}}{2}(\bm{t_i}-\bm{\mu_k})^T(\bm{t_i}-\bm{\mu_k}) \right]
\end{align}

\begin{align}
\frac{\partial Q}{\partial x_0} = \sum_i \sum_k r_{ik} \left[ -2{\sigma_k}^{-1}\frac{\partial \sigma_k}{\partial x_0} - w_i\left( -{\sigma_k}^{-3} \frac{\partial \sigma_k}{\partial x_0}f + \frac{{\sigma_k}^{-2}}{2} \frac{\partial f}{\partial x_0}  \right) \right] 
\end{align}

\begin{align}
\frac{\partial Q}{\partial y_0} = \sum_i \sum_k r_{ik} \left[ -2{\sigma_k}^{-1}\frac{\partial \sigma_k}{\partial y_0} - w_i\left( -{\sigma_k}^{-3} \frac{\partial \sigma_k}{\partial y_0}f + \frac{{\sigma_k}^{-2}}{2} \frac{\partial f}{\partial y_0}  \right) \right] 
\end{align}

\begin{align}
\frac{\partial Q}{\partial z_{w0}} = \sum_i \sum_k r_{ik} \left[ -2{\sigma_k}^{-1}\frac{\partial \sigma_k}{\partial z_{w0}} - w_i\left( -{\sigma_k}^{-3} \frac{\partial \sigma_k}{\partial z_{w0}}f + \frac{{\sigma_k}^{-2}}{2} \frac{\partial f}{\partial z_{w0}}  \right) \right] 
\end{align}

\subsection{Regularized Objective} \label{sec:reg}

Regularized objective function $R$, where $\bm{\mu_k}^0$ is the centroid obtained from the initialization procedure corresponding to event/mirror reflection $k$.

\begin{align}
R = Q - \lambda \sum_{k=0}^K \left \lVert \bm{\mu_k} - \bm{\mu_k}^0 \right \rVert_2^2
\end{align}

\begin{align}
g_k = \left \lVert \bm{\mu_k} - \bm{\mu_k}^0 \right \rVert_2^2
= (\mu_{kx}-\mu_{kx}^0)^2 + (\mu_{ky}-\mu_{ky}^0)^2
= \left(\left(S_2\frac{x_k}{z_{k}}\right)-\mu_{kx}^0\right)^2 + \left(\left(S_2\frac{y_k}{z_{k}}\right)-\mu_{ky}^0\right)^2
\end{align}

\begin{align}
\frac{\partial g_k}{\partial x_0} = 2\left(S_2\frac{x_k}{z_k} - \mu_{kx}^0\right)\left(S_2\frac{\partial}{\partial x_0}\left(\frac{x_k}{z_k}\right)\right) + 2\left(S_2\frac{y_k}{z_k} - \mu_{ky}^0\right)\left(S_2\frac{\partial}{\partial x_0}\left(\frac{y_k}{z_k}\right)\right)
\end{align}

\begin{align}
\frac{\partial g_k}{\partial y_0} = 2\left(S_2\frac{x_k}{z_k} - \mu_{kx}^0\right)\left(S_2\frac{\partial}{\partial y_0}\left(\frac{x_k}{z_k}\right)\right) + 2\left(S_2\frac{y_k}{z_k} - \mu_{ky}^0\right)\left(S_2\frac{\partial}{\partial y_0}\left(\frac{y_k}{z_k}\right)\right)
\end{align}

\begin{align}
\frac{\partial g_k}{\partial z_{w0}} = 2\left(S_2\frac{x_k}{z_k} - \mu_{kx}^0\right)\left(S_2\frac{\partial}{\partial z_{w0}}\left(\frac{x_k}{z_k}\right)\right) + 2\left(S_2\frac{y_k}{z_k} - \mu_{ky}^0\right)\left(S_2\frac{\partial}{\partial z_{w0}}\left(\frac{y_k}{z_k}\right)\right)
\end{align}

\begin{align}
\frac{\partial R}{\partial x_0} =  \frac{\partial Q}{\partial x_0} - \lambda \sum_{k=0}^K \frac{\partial g_k}{\partial x_0}
\end{align}

\begin{align}
\frac{\partial R}{\partial y_0} =  \frac{\partial Q}{\partial y_0} - \lambda \sum_{k=0}^K \frac{\partial g_k}{\partial y_0}
\end{align}

\begin{align}
\frac{\partial R}{\partial z_{w0}} =  \frac{\partial Q}{\partial z_{w0}} - \lambda \sum_{k=0}^K \frac{\partial g_k}{\partial z_{w0}}
\end{align}

\end{document}